\documentclass[aps,prb,twocolumn,reprint,superscriptaddress,floatfix,nofootinbib]{revtex4-2}
\usepackage{siunitx}
\usepackage{physics}  %physics pacakge
\usepackage{graphicx} % Required for inserting images
\usepackage{amssymb}
\usepackage{amsmath} %for equation
\usepackage{soul}  %for highlighting
\usepackage{hyperref}
\usepackage{array}

\newcommand{\PbSe}{Pb$_{1-x}$Sn$_x$Se}

\newcommand{\thepair}{($1$ - $2$)}
\newcommand{\zeroonepair}{($0$ - $1$)}
\newcommand{\onetwopair}{($1$ - $2$)}
\newcommand{\twothreepair}{($2$ - $3$)}
\usepackage{multirow}
\usepackage{threeparttable}
\usepackage{hyperref}
\usepackage[normalem]{ulem}
\hypersetup
{	colorlinks,%
	citecolor=green,%
	linkcolor=red,%
	urlcolor=blue%
}
\allowdisplaybreaks[4]

\usepackage{color}

\newcommand{\notg}{\notag \\}

\begin{document}
\title{Spectral analysis of the magneto-optical response in valley polarized Pb$_{1-x}$Sn$_x$Se}
\author{Xiaoqi Ding}
\thanks{These authors contributed equally to this work.}
\affiliation{Department of Physics, City University of Hong Kong, Kowloon, Hong Kong SAR, China}

\author{Jiashu Wang}
\thanks{These authors contributed equally to this work.}
\affiliation{Department of Physics and Astronomy, University of Notre Dame, Notre Dame, Indiana 46556, USA}
\affiliation{Materials Department, University of California, Santa Barbara, California 93106, USA}

\author{Mykhaylo Ozerov}
\affiliation{National High Magnetic Field Laboratory, Florida State University, 1800 E Paul Dirac Dr, Tallahassee, Florida 32310, USA}

\author{Sara Bey}
\affiliation{Department of Physics and Astronomy, University of Notre Dame, Notre Dame, Indiana 46556, USA}

\author{Muhsin Abdul Karim}
\affiliation{Department of Physics and Astronomy, University of Notre Dame, Notre Dame, Indiana 46556, USA}

\author{\\Seul-Ki Bac}
\affiliation{Department of Physics and Astronomy, University of Notre Dame, Notre Dame, Indiana 46556, USA}

\author{Xinyu Liu}
\affiliation{Department of Physics and Astronomy, University of Notre Dame, Notre Dame, Indiana 46556, USA}

\author{Badih A. Assaf}
\affiliation{Department of Physics and Astronomy, University of Notre Dame, Notre Dame, Indiana 46556, USA}

\author{Yi-Ting Hsu}
\affiliation{Department of Physics and Astronomy, University of Notre Dame, Notre Dame, Indiana 46556, USA}

\author{Xiao Li}
\email{xiao.li@cityu.edu.hk}
\affiliation{Department of Physics, City University of Hong Kong, Kowloon, Hong Kong SAR, China}
\date{\today}

\begin{abstract}
    Since the last century, considerable efforts have been devoted to the study of valley-degenerate narrow gap semiconductors, such as the \PbSe\ alloy. 
     This material possesses band edges at the $L$-points of their Brillouin zone, yielding a valley degeneracy of four. 
    However, in (111)-oriented films, it is still not fully understood how differences between the longitudinal valley, oriented along the growth axis, and the oblique valleys, oriented at an angle with respect to that axis, appear in infrared magneto-optical spectroscopy. 
    In this work, we report a magneto-optical study on this family of alloys, focusing on an anomaly in the interband transition of the absorption strength ratio between longitudinal and oblique valleys under a magnetic field applied along the [111] direction. 
    Based on the Mitchell-Wallis model, we provide a theoretical fit for the experimental transmission data, which quantitatively explains the spectral shape of the data at magnetic fields as high as $\SI{35}{T}$. 
    In particular, we attribute this anomalous absorption strength variation to the carrier density difference between the two types of valleys as well as the field-dependent multiple-beam interference or the Fabry-P\'{e}rot interference. 
    Our analysis also allows for the extraction of the real and imaginary parts of the dielectric function. 
\end{abstract}

\maketitle

\section{Introduction}

The valley degree of freedom in semiconductors and semimetals is behind the realization and prediction of many interesting quantum states. 
In van-der-Waals semiconductors, the spin and valley degrees of freedom are intertwined, resulting in the ability to excite charge carriers from individual valleys with polarized light~\cite{makLightValleyInteractions2018}. 
For instance, in monolayer graphene, quantum isospin ferromagnetism arises when Coulomb interactions spontaneously lift valley degeneracy~\cite{jungLatticeTheoryPseudospin2011, youngSpinValleyQuantum2012}. 
Similarly, in few-layer graphene systems, such as the Bernal bilayer graphene~\cite{zhouIsospinMagnetismSpinpolarized2022}, rhombohedral trilayer graphene~\cite{zhouSuperconductivityRhombohedralTrilayer2021, zhouHalfQuartermetalsRhombohedral2021}, and twisted double bilayer graphene~\cite{liuIsospinCompetitionsValley2022, hsuTopologicalSuperconductivityFerromagnetism2020}, the existence of spin and valley degrees allows for the spontaneous generation of a variety of ground states featuring different spin and valley orderings. 
Finally, in semimetal such as bismuth~\cite{yangPhaseDiagramBismuth2010} and narrow-gap semiconductor like SnTe~\cite{liSUQuantumHall2016}, various valley ordering (such as valley polarized and valley coherent states) is also expected due to the Coulomb interaction. 

While valley ordering due to interactions is attractive, it can also appear in systems where the valley symmetry is explicitly broken due to external perturbations, 
e.g., strain~\cite{krizmanValleyPolarizedQuantumHall2024}, temperature~\cite{bangertMagnetoopticalInvestigationsPhasetransitioninduced1985}, magnetic field~\cite{srivastavaValleyZeemanEffect2015}, etc. 
Therefore, it is crucial to identify the actual mechanism behind the emergence of valley symmetry-breaking states, a task that is by no means straightforward.
For bulk semiconductors, magneto-optical measurements are routinely used to probe the possible existence of various symmetry-breaking states, 
like the ferroelectric state~\cite{bangertMagnetoopticalInvestigationsPhasetransitioninduced1985}, the Volkov-Pankratov state~\cite{bermejo-ortizObservationWeylDirac2023}, the ferromagnetic or the anti-ferromagnetic state~\cite{bacProbingBerryCurvature2025}. 
However, most experiments focused only on identifying the dip positions in infrared magneto-optical spectra, which directly relate to the selection rules governing optical transitions. 
The question of whether analyzing the absorption strength or dip intensity can help identify symmetry-breaking mechanisms remains both important and largely unexplored.

In this work, we study this crucial question in the IV-VI material \PbSe, which is a narrow-gap semiconductor that hosts band edges at four $L$-points and offers a platform to study the valley degree of freedom as the electronic band structure is tuned. 
Historically, this family of materials was known as ideal platforms for studying fundamental properties of narrow-gap bulk semiconductors, such as thermoelectric performance and optoelectronics~\cite{mitchellTheoreticalEnergyBandParameters1966,Martinez1973,Martinez1973a,Martinez1973b,Martinez1975}. 
Moreover, it has been known that their band inversions can be modulated by continuously varying the alloy composition $x$~\cite{straussInversionConductionValence1967}. 
More recently, this family of materials is better known for their connections to topological phases of matter. 
In fact, the observed band inversion is now understood to be accompanied by a quantum phase transition from a narrow gap semiconductor to a topological crystalline insulator (TCI) as the alloy composition $x$ goes above $0.16$~\cite{fuTopologicalCrystallineInsulators2011b,hsiehTopologicalCrystallineInsulators2012,dziawaTopologicalCrystallineInsulator2012,krizmanDiracParametersTopological2018,tikuisisLandauLevelSpectroscopy2021}. 
Unlike topological insulators protected by the time-reversal symmetry~\cite{kaneQuantumSpinHall2005a, bernevigQuantumSpinHall2006, hasanColloquiumTopologicalInsulators2010}, TCIs are protected by crystal point group symmetry, such as discrete rotational symmetry~\cite{fuTopologicalCrystallineInsulators2011b} and mirror symmetry~\cite{hsiehTopologicalCrystallineInsulators2012}. 
Furthermore, the metallic surface states of TCIs only appear on surfaces that respect the crystalline symmetry.

The \PbSe\ compounds were the first reported TCI~\cite{hsiehTopologicalCrystallineInsulators2012, fuTopologicalCrystallineInsulators2011b} and were later studied as a model system to elucidate the properties of mirror-symmetry protected topological surface states~\cite{dziawaTopologicalCrystallineInsulator2012, okadaObservationDiracNode2013}. 
Additionally, varying temperatures and alloy compositions allow one to tune not only the energy gap but also the band velocity and the Fermi surface anisotropy of these materials~\cite{krizmanDiracParametersTopological2018}.
However, the interest in
 quantum Hall and Landau-level physics in IV-VI materials precede the era of topological phases of matter. 
The high mobility of PbTe and other IV-VI materials allowed a thorough mapping of their low-energy band structure parameters through Landau level spectroscopy~\cite{ 
calawaMagneticFieldDependence1969, burkhardBandpopulationEffectsIntraband1979,
bauerMagnetoopticalPropertiesIV1980,
assafMagnetoopticalDeterminationTopological2017, krizmanDiracParametersTopological2018}.

\begin{figure*}[!]
    \includegraphics[width=\textwidth]{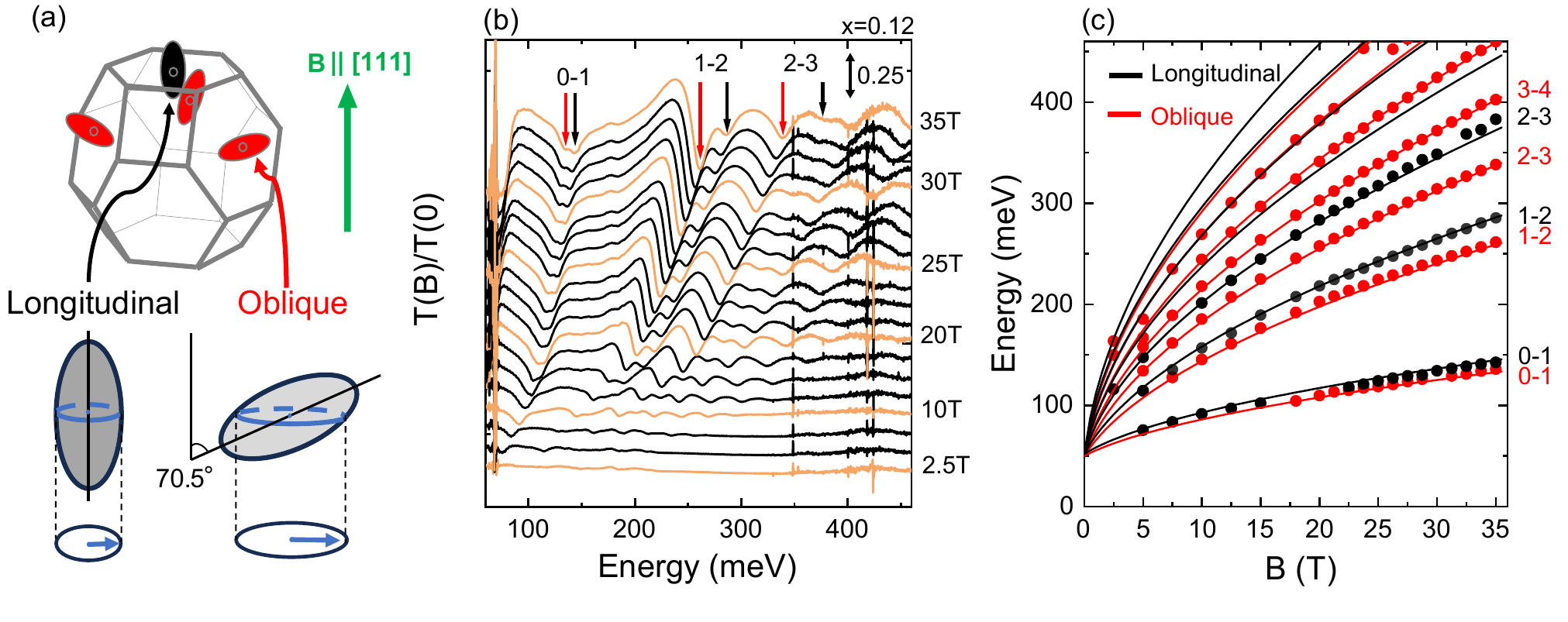}
    \caption{(a) Sketch of the oblique and longitudinal valleys of \PbSe. 
    (b) Relative transmission $T(B)/T(0)$ for Sample S1 with $x=0.12$. 
    Transitions from different valleys are marked by red (oblique) and black (longitudinal) arrows. 
    Labeled spectra for specific fields are plotted in orange for clarity. 
    The curves have been shifted for clarity.
    (c) Landau level fan chart extracted from the spectra in (b) and fitted by Eq.~\eqref{Eq:1}. Interband transitions between Landau levels $N$ and $N+1$ are labeled on the right.}
    \label{fig:1}
\end{figure*}

The band edges of \PbSe\ are located at the $L$-points of the Brillouin zone. 
The 3D Fermi surface of these materials consists of four ellipsoids with their great axis oriented along $\langle 111 \rangle$ axis, as shown in Fig.~\ref{fig:1}(a). 
In (111)-oriented layers, Fermi surface anisotropy yields two types of valleys. 
Among them, the valley aligned with the growth axis (the [111] direction) is known as the longitudinal valley, while the other three valleys, oriented at an angle with the growth axis, are known as the oblique valleys. 
The Landau levels of bulk IV-VI semiconductors have been extensively studied using infrared magneto-optical spectroscopy (IRMS). This technique is an established tool to extract the band parameters of electrons in solids by probing optical transitions between Landau levels~\cite{ 
maltzMagnetoreflectionStudiesBismuth1970,
orlitaObservationThreedimensionalMassless2014, 
krizmanDiracParametersTopological2018,
jiangUnravelingTopologicalPhase2020}. 
However, information pertaining to the spectral shape of infrared magneto-optical spectra has been overlooked in most experiments.

The focus of Ref.~\cite{tikuisisLandauLevelSpectroscopy2021} is to identify the surface states of \PbSe\ samples after the band inversion with the alloy composition $x>0.16$. In contrast,
the present paper focuses on investigating a ubiquitous spectral anomaly~\cite{krizmanDiracParametersTopological2018,assafMagnetoopticalDeterminationTopological2017,phuphachongMagneto-spectroscopy2017} before the band inversion with $x<0.16$. 
This anomaly appears in the relative absorption strength of magneto-optical transitions between the longitudinal and oblique valleys.
Despite the expected three-to-one degeneracy ratio, our experimental observation shows that the optical transitions between low-index Landau levels are shown to violate this ratio. 
Such a spectral anomaly has been observed in the same family of materials in previous experiments but never studied. 
To resolve this puzzle, we develop a theoretical model to capture the spectral response of the Landau level transitions.  
By applying this method to analyze our experimental data obtained for (111)-oriented \PbSe\  epilayers, we find that the anomalous intensity ratio arises from two important factors: the carrier density difference between the two types of valleys (i.e., valley polarization) and the field-dependent multiple-beam interference effects. 
When included, the model accurately reproduces our data, allowing the extraction of a disorder-broadening parameter and yielding the relative carrier population for each valley. 
Through this analysis, the real and imaginary parts of the dielectric function are also extracted. 
Our theoretical approach provides a comprehensive framework for understanding the anomalous absorption strength in infrared magneto-optical spectroscopy of IV-VI semiconductors and elucidates the actual mechanism behind valley ordering in our sample.

The structure of the paper is the following:
In Section~\ref{Sec:Model Parameter}, we build up an effective Hamiltonian and extract the model parameters from the Landau fan diagram obtained from the experiment.
In Section~\ref{Sec:Anomalous Ratio}, we analyze the infrared magneto-optical spectroscopy data and explain the observed anomalous absorption strength ratio variations. 
In Section~\ref{Sec:Conclusion}, we provide additional discussions on the results and conclude the paper.

\section{The Landau fan diagram \label{Sec:Model Parameter}}

Experimentally, we carry out IRMS measurements using infrared setups at the National High Magnetic Field Laboratory (NHMFL).
The in-house developed vacuum FT-IR spectrometer was used in conjunction with a resistive magnet, enabling transmission measurements at magnetic fields up to $B = \SI{35}{T}$ and temperatures as low as $T=\SI{5}{K}$. 
Infrared radiation emitted from the spectrometer's Globar source propagates inside an evacuated optical beamline to the top of the cryostat, then entering the probe's light pipe and continuing down to the sample. 
The measurement is set up in Faraday geometry~\cite{wangEnergyGapTopological2023}, 
and the transmission signal is collected by a silicon bolometer positioned a short distance below the sample.
The field-independent spectral features were removed by normalizing the transmission spectrum at each magnetic field to the spectrum measured at zero magnetic field.
These high-field infrared data were additionally supplemented by a dataset with a significantly better signal-to-noise ratio, obtained using a different  FT-IR setup coupled with a \SI{17.5}{T} superconducting magnet.
In this work, we report measurements on two samples, S1 and S2, in Table~\ref{Tab:parameters_S1}, both of which are grown on BaF$_2$ (111) substrate by molecular beam epitaxy.

Figure~\ref{fig:1}(b) shows infrared magneto-optical spectra taken on \PbSe\ with $x=0.12$ (Sample S1). 
By dividing the transmission spectra $T(B)$ by the spectra measured at zero magnetic field $T(0)$, we extract the relative transmission that contains only field-dependent transitions. 
The relative transmission exhibits a set of minima that shift to higher energies with increasing magnetic field originating from interband transitions between the Landau levels of the conduction and valence band. 
From the spectra, we can construct a Landau level fan chart shown in Fig.~\ref{fig:1}(c), where the energy of dips in Fig.~\ref{fig:1}(b) is re-plotted as a function of the magnetic field. 
Note that water cooling in the resistive setup generates additional noise at the high photon energy range in the spectra.

\subsection{The $k\cdot p$ model for \PbSe} 
To understand the observed Landau fan diagram, we theoretically introduce the  $k\cdot p$ Hamiltonian for \PbSe\ near the four $L$-points~\cite{mitchellTheoreticalEnergyBandParameters1966,adlerModelMagneticEnergy1973,bauerMagnetoopticalPropertiesSemimagnetic1992,assafNegativeLongitudinalMagnetoresistance2017, krizmanInteractionInterfaceMassive2022}, 
\begin{align} \label{Eq:k_dot_p_model}
    &\mathcal{H} (\vb{k}) = H(\vb{k}) + \Lambda \cdot \mathbb{I}, \notg
    &H(\vb{k})  = \mqty{ 
        \mqty[
        -\varepsilon_k & 0 &  \hbar v_z k_z &  \hbar v_c k_{-} \\
        0 & -\varepsilon_k &  \hbar v_c k_{+} & -\hbar v_z k_z \\
        \hbar v_z k_z & \hbar v_c k_{-} & \varepsilon_k & 0 \\
         \hbar v_c k_{+} & -\hbar v_z k_z & 0 & \varepsilon_k
    ] }. 
\end{align}
In the above equation, $\varepsilon_k \equiv \Delta + \hbar^2 k_{\perp}^2/2 m$, $v_{z}$ is the velocity along the $z$-axis (which is parallel to the [111] direction), $k_\pm=k_x \pm i k_y$, $k_{\perp}^2=k_x^2 + k_y^2$, $v_c$ is the velocity in the $x$-$y$ plane,  
and $m$ represents the contribution from the far bands to the effective mass. 
We ignored a potential $k_z^2$ term in $\varepsilon_k$, as its contribution is much smaller than the other terms~\cite{krizmanInteractionInterfaceMassive2022}. 
The symbol $\Lambda$ labels a relative shift of the valley splitting and equals $0$ $(\Delta_S)$ for the longitudinal (oblique) valley.

In the presence of a strong perpendicular magnetic field, Landau levels appear in the energy spectra. 
After applying the Peierls substitutions and 
some simplifications~\cite{assafNegativeLongitudinalMagnetoresistance2017}, 
we can write the the Hamiltonian $H(\vb{k})$ as
\begin{widetext}
    \begin{align}   \label{Eq:spin_splitting_Hamiltonian}
        H_{B}(N, k_z)=
        \mqty{ 
            \mqty[
            -\Delta - \hbar \omega_S N_{-}  & 0 &  \hbar v_z k_z &  \hbar \omega_R \sqrt{N} \\
            0 & -\Delta - \hbar \omega_S N_{+}  &  \hbar \omega_R \sqrt{N} & -\hbar v_z k_z \\
            \hbar v_z k_z & \hbar \omega_R \sqrt{N} & \Delta + \hbar \omega_S N_{-}  & 0 \\
            \hbar \omega_R \sqrt{N} & -\hbar v_z k_z & 0 & \Delta + \hbar \omega_S N_{+}
        ] }. 
    \end{align} 
\end{widetext}
The above Hamiltonian is written in the basis $\mqty[V^{\frac{1}{2}} \ket{N-1}, V^{-\frac{1}{2}} \ket{N}, C^{\frac{1}{2}} \ket{N-1}, C^{-\frac{1}{2}} \ket{N}]$, where $V$ and $C$ denote the valence and conduction bands, respectively, and the superscripts $\pm 1/2$ denote the two eigenstates that form a Kramers pair in this subspace. 
In the above Hamiltonian, $N$ is the Landau level index, $N_{\pm} \equiv N \pm 1$, and $\Delta$ is half of the band gap. 
We have also introduced the relativistic energy $\hbar \omega_R = \sqrt{2 \hbar e B v_c^2}$ and the non-relativistic energy $\hbar \omega_S = \hbar e B / m$.  

The eigenvalues of the effective Hamiltonian are
\begin{align} 
    \begin{cases}
        E_{\lambda=\pm, N, \xi=\pm} (k_z)   = \lambda \sqrt{P_{N, \xi}^2 + \hbar^2 v_z^2 k_z^2} + \Lambda,  &   N>0 \\
        E_{\lambda=\pm, 0, \xi=+} (k_z) = \lambda \sqrt{P_{0, \xi=+}^2 + \hbar^2 v_z^2 k_z^2} + \Lambda, &  N=0 
    \end{cases},
\label{Eq:energy_eigenvalues}
\end{align}
where
\begin{equation}
    P_{N, \xi=\pm} = \xi \hbar \omega_S + \sqrt{N \hbar^2 \omega_R^2 + (\Delta + N \hbar \omega_S)^2} . 
\end{equation} 
Here, $\lambda = \pm 1$ labels the conduction and valence band, respectively, and $\xi = \pm 1$ is the index that represents the Kramers pair $s=\mp 1/2$. 
The corresponding eigenstates will be denoted as $\ket{\lambda, N, \xi, k_z}$. 

\subsection{Extracting model parameters}

\begin{figure} [t]
    \centering
    \includegraphics[width=\columnwidth]{"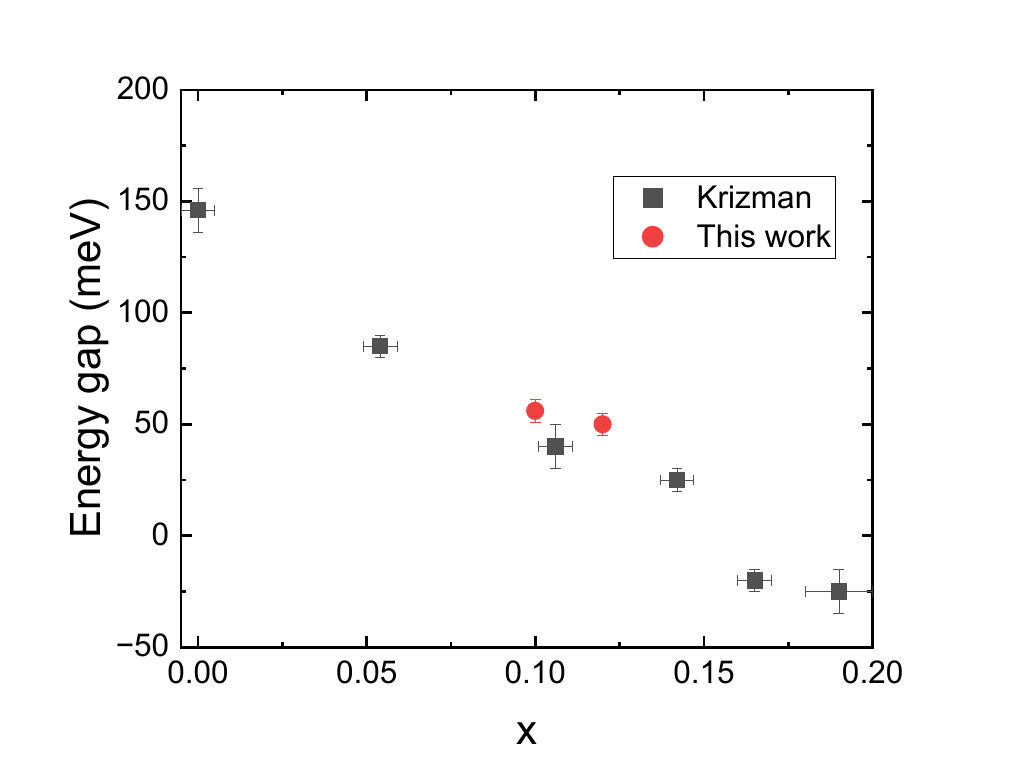"}
    \caption{Energy gap as a function of the alloy composition $x$. Our data points lie on
    the upper end of the error bars determined by Krizman et al.~\cite{krizmanDiracParametersTopological2018}.
    \label{Fig:compare_with_refs}}
\end{figure}

The above $k\cdot p$  model can now be used to understand the Landau fan chart in Fig.~\ref{fig:1}(c). 
In particular, each dot in the fan chart corresponds to the dip of an interband transition characterized by the divergent DOS at $k_z = 0$ and the selection rule $\Delta{N} = \pm1$ and  $\Delta s=\pm1$. 
Moreover, the following interband transitions have the same dip energy at $k_z=0$: 
\begin{align} \label{Eq:interband_transition}
             \text{(i)}    \     \ket{\lambda = -, N, \xi = -} \rightarrow \ket{\lambda = +, N+1, \xi = +}, \notg
             \text{(ii)}    \       \ket{\lambda = -, N, \xi = +} \rightarrow \ket{\lambda = +, N+1, \xi = -}, \notg
             \text{(iii)}   \       \ket{\lambda = -, N+1, \xi = -} \rightarrow \ket{\lambda = +, N, \xi = +}, \notg
             \text{(iv)}   \       \ket{\lambda = -, N+1, \xi = +} \rightarrow \ket{\lambda = +, N, \xi = -}. 
\end{align}
Later on, without explicitly mentioning, we use the label $\ket{\lambda, N, \xi} \equiv \ket{\lambda, N, \xi, k_z=0}$. As a result, we will use the symbol ($N$ - $N$+$1$) to denote the collection of these transitions. 
Such considerations yield the following energy for the ($N$ - $N$+$1$) interband transition~\cite{wangFactorTopologicalInterface2023, wangEnergyGapTopological2023}:
\begin{align} 
\label{Eq:1}
\Delta E          &= \sqrt{(\Delta+ (N+1) \hbar \omega_S )^2 + (N+1) \hbar^2 \omega_R^2} 	\notg
			     &+ \sqrt{(\Delta+N \hbar \omega_S )^2 + N \hbar^2 \omega_R^2} ,
\end{align}
which are plotted as solid lines in Fig.~\ref{fig:1}(c). 
By fitting the solid lines with the experimental data, we obtain the model parameters in Table~\ref{Tab:parameters_S1}. 
A comparison between our data and Krizman et al.~\cite{krizmanDiracParametersTopological2018} is shown in
Fig.~\ref{Fig:compare_with_refs}. 
Our data points lie on the upper end of the error bars determined by theirs, but are largely consistent.

\begin{table*}[!]
    \centering
    \caption{\label{Tab:parameters_S1}Parameters of the two samples. }
    \begin{threeparttable}
        \begin{tabular}{c|c|c|c|c|c|c|c|c|c}
        \hline\hline
        \multirow{2}{*} & \multirow{2}{*}{Valley}        & \multirow{2}{*}{$\Delta$ (meV)}  & \multirow{2}{*}{$v_c(\SI{d5}{m/s})$}  & \multirow{2}{*}{$v_z \tnote{*}\ \  (\SI{d5}{m/s})$}   & \multirow{2}{*}{$m$($m_e$)} & \multirow{2}{*}{$\Delta_S \tnote {$\dagger$} $ (meV)} &$p$-type carrier density for   & \multirow{2}{*}{$g_v$}  & \multirow{2}{*}{$d$ (nm)}  \\  
        \multirow{2}{*}{}&          &   &   &   &  &  & each individual valley \tnote{$\dagger$} $(\SI{d17}{cm^{-3}}$)   &     &           \\ \hline
            \multirow{2}{*}{S1  ($x=0.12$)}&L    & 25  & 5.20  & 4.11  & 0.25 & \multirow{2}{*}{28.50} & 0.52  & 1    & \multirow{2}{*}{518}           \\ \cline{2-6} \cline{8-9}
            \multirow{2}{*}{}& O         & 25  & 4.70  & 5.14  & 0.30 &  &  3.2  & 3    &           \\ \hline
            \multirow{2}{*}{S2  ($x=0.10$)}&L    & 28  & 5.38  & 4.19  & 0.40 & \multirow{2}{*}{11.46} & 1.1  & 1   & \multirow{2}{*}{990}           \\ \cline{2-6} \cline{8-9} 
            \multirow{2}{*}{}& O         & 28  & 4.83  & 5.32  & 0.40 &   &  2.2 & 3    &          \\ \hline \hline
        \end{tabular}
        \begin{tablenotes}
            \footnotesize 
            \item[*] $v_z$ is derived from the $x$-$y$ plane velocity $v_c$ of longitudinal and oblique valleys~\cite{krizmanDiracParametersTopological2018}.
            \item[$\dagger$] These quantities are inferred from Appendix~\ref{Appendix:Determining Carrier Density}.
        \end{tablenotes}
    \end{threeparttable}
\end{table*}

\begin{figure} [b]
    \centering
    \includegraphics[width=0.9\columnwidth]{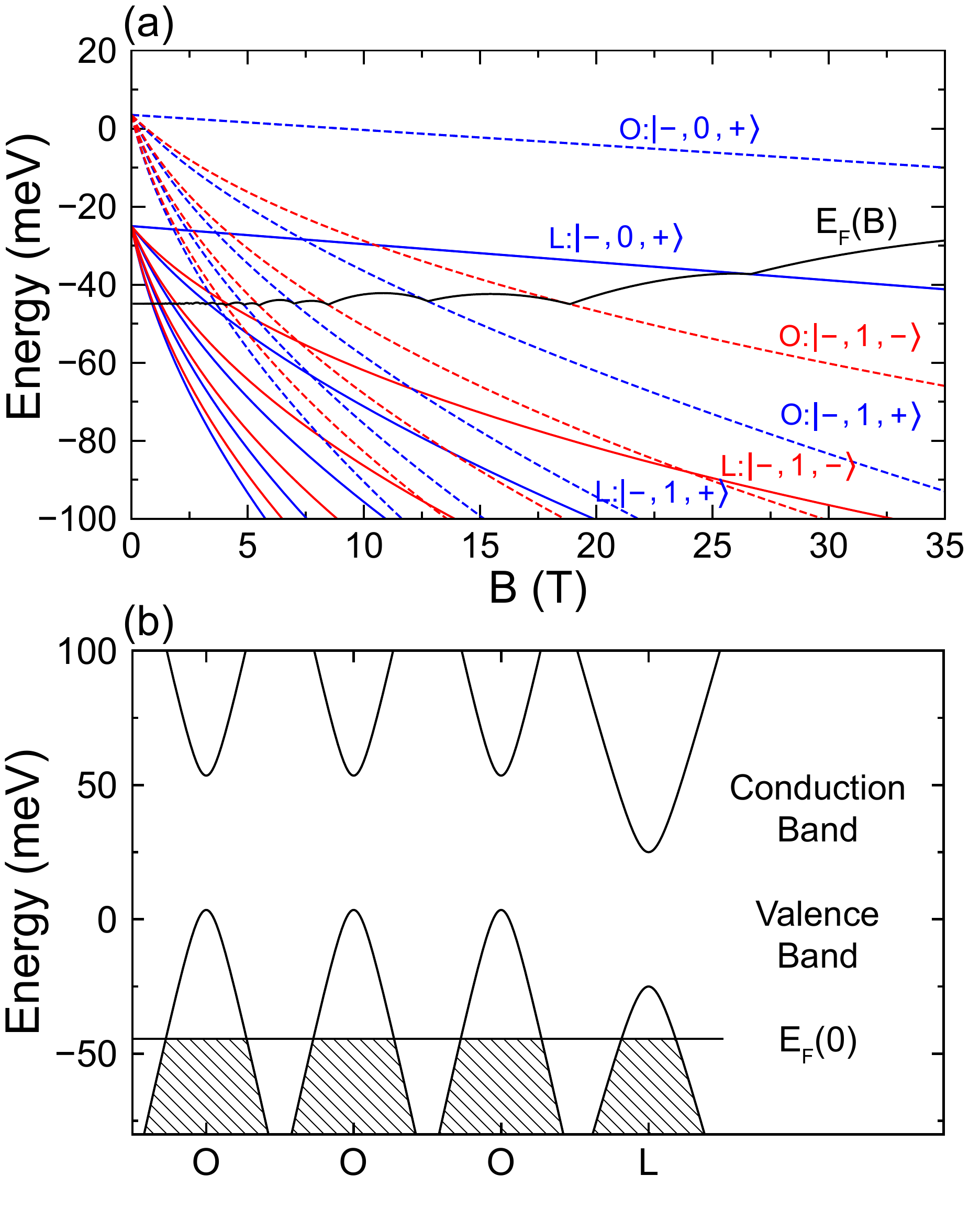}    
    \caption{\label{Fig:Fermi_level}
    Landau fan diagram of the Sample S1. 
    (a) Solid and dashed lines show the Landau levels of longitudinal and oblique valleys, respectively. 
    In particular, we use $\ket{\lambda, N, \xi}$ to label each Landau level at $k_z=0$. 
    The blue (red) lines illustrate the Landau level with the index $\xi=+1$ ($-1$). 
    Moreover, the black line in the figure marks the position of the Fermi level.  
    (b) Energy dispersion along $k_z$ at $B=\SI{0}{T}$ with $k_x=k_y=0$. The valley splitting results in the different populations of holes (empty states below the gap) in different valleys.
    }
\end{figure}

The infrared magneto-optical spectra also allow us to extract the Fermi energy $E_F$ of the system. 
Specifically, in the presence of a strong magnetic field, the $p$-type carrier density of the system can be estimated by
\begin{align}	\label{Eq:total_carrier_density}
    &n_\text{total}(E_F)  = \frac{g_{L}}{S} \sum_\text{valleys} \sum_{N,\xi} \int_{-\infty}^{\infty} \qty[1-f(E_{\lambda, N, \xi}(k_z))] \frac{d k_z}{2 \pi} \notg
            &= \frac{1}{2 \pi^2 l_B^2} \frac{1}{\hbar v_z} \sum_\text{valleys} \sum_{N, \xi} \sqrt{(E_F')^2 - P_{N, \xi}^2} \Theta(-\qty|P_{N, \xi}| - E_F'), 
\end{align}
where $E_F' = E_F-\Lambda$, $g_{L} = S/2 \pi l_B^2$ is the degeneracy of each Landau level with $l_B = \sqrt{\hbar/eB}$ being the magnetic length, and $S$ is the cross-section area of the sample perpendicular to the magnetic field. 
Moreover, the Fermi-Dirac distribution $f(E_{\lambda, N, \xi}(k_z))$ is taken to be a step function in the zero-temperature limit and $\lambda=-1$ represents the valence band. 
The details of how we determine the carrier density are provided in Appendix~\ref{Appendix:Determining Carrier Density}.
After obtaining the carrier density of the system, the Fermi level can be uniquely determined from Eq.~\eqref{Eq:total_carrier_density} at various magnetic fields.
The variation of the Fermi level as a function of the magnetic field is shown in Fig.~\ref{Fig:Fermi_level}(a). 
It is observed that the longitudinal valley enters the quantum limit (where only the lowest Landau level is unoccupied) at a significantly lower magnetic field compared to the oblique valleys.
As the magnetic field continues to increase, 
the carrier density of the longitudinal valley becomes completely depleted at around \SI{26.6}{T} as shown in Appendix~\ref{Appendix:Determining Carrier Density}. 
We also observe the population transfer between two valleys, a phenomenon that has been reported in the literature~\cite{bauerMagnetoopticalPropertiesIV1980, burkhardBandpopulationEffectsIntraband1979, adlerModelMagneticEnergy1973}. 
Notably, we find that a finite valley splitting 
$\Delta_S$ between the two valleys is necessary to fit the experimental data (see Appendix~\ref{Appendix:Determining Carrier Density} for further details).
Meanwhile, even in micron-thick epilayers of IV-VI materials, the uneven valley populations could arise from the thermal strain resulting from the thermal expansion coefficient mismatch between the substrate and the material~\cite{burkeMagnetophononShubnikovde1978,assafMassiveMassless2016}. 
Such a thermal strain is known to cause a splitting between the longitudinal and oblique valleys.  
A similar splitting resulting from lattice strain has recently enabled the observation of a valley-polarized quantum Hall effect in IV-VI quantum wells~\cite{krizmanValleyPolarizedQuantumHall2024}.

\begin{figure*}[!]
    \centering
    \includegraphics[width=\textwidth]{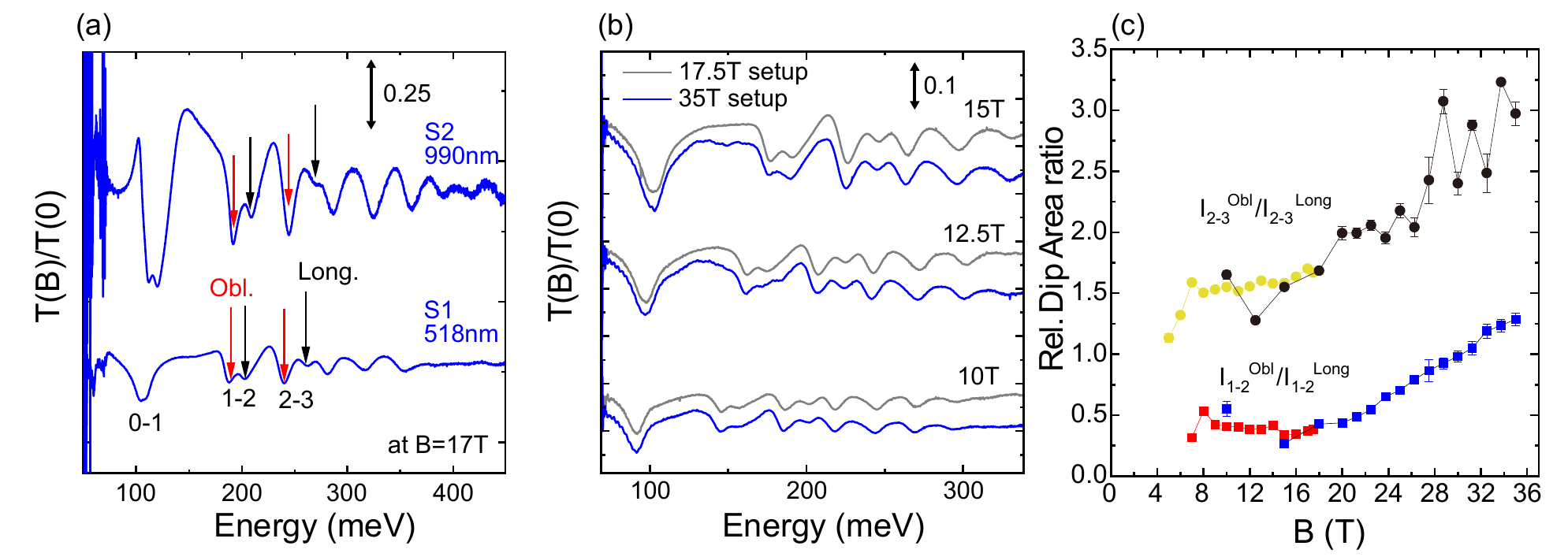}
    \caption{(a) Comparison of the spectra of the experiment acquired at 17T for samples S1 and S2. 
    (b) Comparison between the spectra obtained at $\SI{10}{T}$, $\SI{12.5}{T}$, $\SI{15}{T}$ and using two different IRMS setups for the sample S1. The spectra between the two setups are manually shifted for better shape comparison. 
    (c) Dip area ratio between the oblique and longitudinal valley. The red and blue data points (measured in different setups) correspond to the \onetwopair\ interband transition, and the yellow and black data points correspond to the \twothreepair\ interband transition.}
    \label{fig:3}
\end{figure*}

\section{Anomalous Absorption Strength Ratio in the Infrared magneto-optical Spectra \label{Sec:Anomalous Ratio}}

\begin{figure*}[t]
    \centering
    \includegraphics[width=0.9\textwidth]{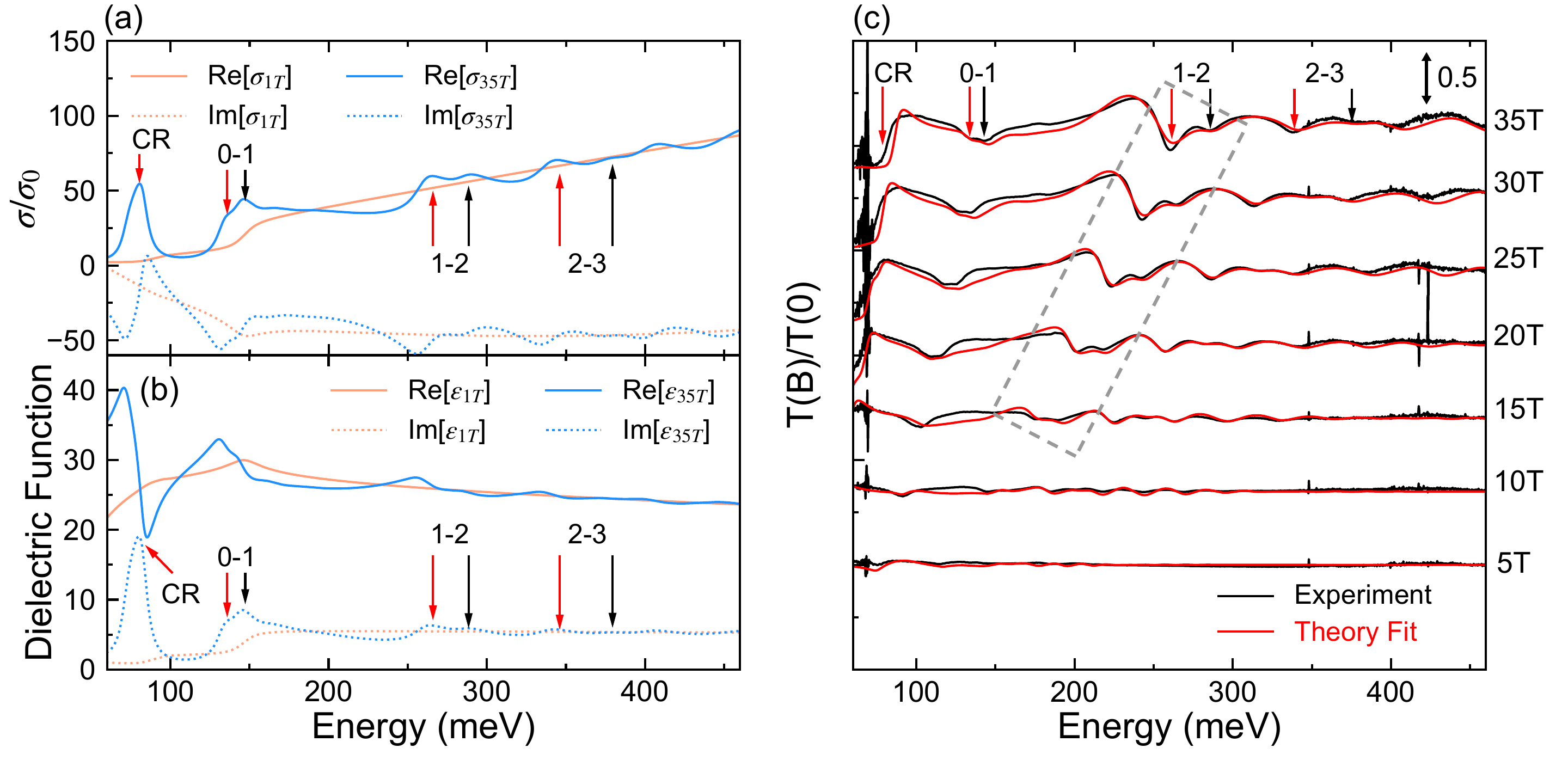}
    \caption{ \label{Fig:total_spectrum_epsilon_S1} 
     The theoretically calculated real and imaginary part of the (a) conductivity $\sigma$ and (b) dielectric function $\varepsilon$  at $B=\SI{1}{T}$ and $B=\SI{35}{T}$. These results are obtained by averaging over left-handed and right-handed circularly polarized light. The unit conductivity $\sigma_0=e^2/8 \pi\hbar l_B$ is calculated at $B=\SI{1}{T}$.
    (c) The infrared magneto-optical spectra of the sample S1. The black and red lines correspond to the experiment and theory, respectively. 
   The dashed box highlights the anomalous absorption strength variations of the \thepair\ interband transition.}
\end{figure*}

One of the most interesting features in the experimental data is the anomalous absorption strength ratio variations, which we discuss in this section.  

\subsection{Absorption strength ratio variations in the experiment}

In Fig.~\ref{fig:1}(b), we have marked the \zeroonepair, \onetwopair\, and \twothreepair\ interband transition dips according to Fig.~\ref{fig:1}(c). 
Dips from the intraband cyclotron resonances (CR) may be present at lower energies at $\SI{35}{T}$, which quickly merges into the noise region and cannot be extracted reliably. 
Transitions corresponding to oblique and longitudinal valleys are marked by red and black arrows, respectively. 
Because the oblique valleys are three-fold degenerate ($g_v=3$) and the longitudinal valley is non-degenerate ($g_v=1$), one would have naively expected that the absorption strength of magneto-optical transitions from the oblique valley to be close to three times stronger than those from the longitudinal valley. 
However, it is evident from Fig.~\ref{fig:1}(b) that while the \twothreepair\ interband transition roughly follows this expectation, the \onetwopair\ interband transition does not. 
The latter transition violates this ratio over the entire magnetic field range.

We can confirm that this anomalous absorption strength ratio variation is an intrinsic property of the material. 
For example, Fig.~\ref{fig:3}(a) presents infrared magneto-optical spectra measured for another sample S2 at $\SI{17}{T}$. 
The anomaly seen in Sample S1 for the \onetwopair\ interband transition is clearly reproduced in this second sample. 
Additionally, Fig.~\ref{fig:3}(b) demonstrates that the reproducibility of the spectral shape seen in Fig.~\ref{fig:1}(b) for measurements taken on two different setups at the NHMFL (a $\SI{35}{T}$ resistive magnet and a $\SI{17.5}{T}$ superconductor coil). 

Figure~\ref{fig:3}(c) plots the ratio between the integrated absorption strength, or integrated dip intensity, of the oblique and the longitudinal valley transitions for the \onetwopair\ and \twothreepair\ interband transitions. 
The integrated absorption strength is extracted by performing a Gaussian fit and taking the integral over the fitted function in a narrow range of energies spanning each transition. We should note that the uncertainty of such a fit is influenced by the definition of the baseline of the spectra. 
Thus, it only serves as a means of extracting the changing trend of the ratio versus magnetic fields and does not represent the value of the real ratio. 
However, it is evident from this fit that the integrated absorption strength ratio for the \onetwopair\ interband transition is significantly smaller than that found for the \twothreepair\ interband transition.

\subsection{Optical conductivity and transmission}

In order to understand the observed absorption strength ratio variations, we need to calculate the theoretical transmission spectra based on the model in Eq.~\eqref{Eq:spin_splitting_Hamiltonian}. 
In general, the transmission properties of a bulk semiconductor are related to its dielectric function, which can be written generally as~\cite{dresselhausSolidStateProperties2018} 
\begin{equation} 
    \varepsilon = \varepsilon_\text{core} + i \frac{\sigma}{\omega \varepsilon_0}, \label{Eq:epsilon}
\end{equation}
where $\sigma$ is the (complex) conductivity of the material, $\omega$ is the frequency of the incident light, and $\varepsilon_0$ is the vacuum permittivity.
$\varepsilon_\text{core}$ represents the contributions from far bands outside our energy window.
In the literature, the value of $\varepsilon_\text{core}$ of our energy window typically ranges from $19.8$ to $21.5$ for PbSe~\cite{suzukiOpticalPropertiesPbSe1995, shaniCalculationRefractiveIndexes1985, pascherMagnetooptical1988, albanesiCalculatedOpticalSpectra2005, falkovskyOpticalPropertiesGraphene2008, tikuisisLandauLevelSpectroscopy2021}.
In this work, we take  $\varepsilon_\text{core}=21$, with which we can produce a refraction index $\bar{n}$ (see Appendix~\ref{Appendix:OpticalConstants}) consistent with values reported in the literature~\cite{shaniCalculationRefractiveIndexes1985, suzukiOpticalPropertiesPbSe1995, cardonaOpticalPropertiesBand1964b}.

Based on the Kubo formula, the dynamical conductivity for the right-handed ($+$) circularly polarized light and the left-handed ($-$) circularly polarized light can be written as
\begin{equation}
\label{Eq:conductivity}
    \sigma_{\pm} (\omega) = \frac{g_{L} g_v e^2 \hbar}{i \mathcal{V}} \sum_{\alpha, \beta} 
    \frac{f(E_\alpha) - f(E_\beta)}{E_\alpha - E_\beta} \frac{\qty|\mel{\beta}{\hat{v}_\pm}{\alpha}|^2}{E_\alpha - E_\beta + \hbar \omega + i \Gamma},
\end{equation}
where $\mathcal{V}$ is the volume of the sample, $g_v=1$ and $g_v=3$ are the degeneracy of the longitudinal and oblique valley, respectively, and $\omega$ is the frequency of incident light. 
In addition, $\alpha$ and $\beta$ label all eigenstates $\ket{\lambda,N,\xi,k_z}$.
The velocity operator $\hat{v}_{\pm} = (\hat{v}_x \pm i \hat{v}_y)/\sqrt{2}$ are defined as $\hat{v}_x = \pdv*{H}{p_x}$, $\hat{v}_y = \pdv*{H}{p_y}$. 
Finally, $\Gamma$ is a broadening parameter. 
When we fit the experimental data with our formula, we found that the broadening parameter $\Gamma$ is best taken to be $\Gamma(\hbar \omega)=a+b\hbar \omega$~\cite{tikuisisLandauLevelSpectroscopy2021}, where $a$ and $b$ are fitting parameters. 
Further details for using this dynamic broadening model are provided in Appendix~\ref{Appendix:Numerics}.

In Fig.~\ref{Fig:total_spectrum_epsilon_S1}(a), we present the numerically calculated conductivity of the sample at two different magnetic fields of $B = \SI{1}{T}$ and $B = \SI{35}{T}$. 
At $B = \SI{1}{T}$, the real part of the conductivity $\Re[\sigma(\omega)]$ has two interesting features. 
First, it shows a sudden increase for $\hbar\omega \sim \SI{150}{meV}$, which is due to the absorption edge. 
Second, it approaches a linear function of the photon energy in the high-frequency limit. 
At $B = \SI{35}{T}$, the whole curve starts to oscillate around the $B=\SI{1}{T}$ case. 
Note that we label the intraband cyclotron resonance as CR in the figure, which mainly arises from the intraband transitions between the $N=0$ and $N=1$ Landau levels in the valence band. 

Next, we consider the dielectric function shown in Fig.~\ref{Fig:total_spectrum_epsilon_S1}(b), which inherits many features from the conductivity. 
For example, because $\Im[\varepsilon] = \Re[\sigma]/\omega\varepsilon_0$, the asymptotic behavior of $\Re[\sigma]\sim\omega$ produces a constant $\Im[\varepsilon]=5.5$. 
Regarding the energy window concerned, with $\varepsilon_\text{core} = 21$, the real part of the calculated dielectric function $\Re[\varepsilon]$ is close to 25 as reported in the literature~\cite{suzukiOpticalPropertiesPbSe1995,albanesiCalculatedOpticalSpectra2005, falkovskyOpticalPropertiesGraphene2008}.
Meanwhile, we introduce two important optical constants, the refractive index $\bar{n}$ and the optical extinction coefficient $k$. They are related to the dielectric function $\varepsilon$ as
\begin{equation}	
    \bar{n} = \sqrt{\frac{\qty|\varepsilon| + \Re[\varepsilon]}{2}}, \quad k= \sqrt{\frac{\qty|\varepsilon| - \Re[\varepsilon]}{2}}. \label{Eq:refractive_index}
\end{equation}
In the high-frequency regime, as shown in Appendix~\ref{Appendix:OpticalConstants}, the refractive index exhibits $\bar{n} \approx 5$.
To see this, note that $\Re[\varepsilon]^2 \gg \Im[\varepsilon]^2$, we can estimate that the refractive index $\bar{n} \approx \sqrt{\Re[\varepsilon]} \approx 5$, which is consistent with the 
value in the literature~\cite{shaniCalculationRefractiveIndexes1985, suzukiOpticalPropertiesPbSe1995, cardonaOpticalPropertiesBand1964b}.

With the dielectric function at hand, we can finally obtain the slab transmission as~\cite{palikInfraredMicrowaveMagnetoplasma1970, heavensOptical1960}
\begin{equation}   \label{Eq:slab_transmission}
    T(B)  = e^{-Ad} \frac{(1 - \qty|r|^2)^2 + 4\qty|r|^2\sin^2(\phi_r)}{(1 - \qty|r|^2 e^{-A d})^2 + 4\qty|r|^2 e^{-A d} \sin^2(\alpha d + \phi_r)}, 
\end{equation}
where $d$ is the thickness of the sample, and all other parameters are functions of $\bar{n}$ and $k$. 
A brief derivation of Eq.~\eqref{Eq:slab_transmission} is given in Appendix~\ref{Appendix:Slab Transmission}. 
Specifically, $\alpha = \omega \bar{n}/c$, and $A = 2\omega k/c$ is the absorption coefficient. 
In addition, $\qty|r|$, $\phi_r$ in Eq.~\eqref{Eq:slab_transmission} are the amplitude and phase of the half-space reflection coefficient $r$, respectively, defined as 
\begin{equation} \label{Eq:half_space_reflection}
    r  = \frac{1-(\bar{n} + i k)}{1+(\bar{n} + i k)}. 
\end{equation}
The results of these optical coefficients are discussed in Appendix~\ref{Appendix:OpticalConstants}.

By combining the above Eq.~\eqref{Eq:epsilon}-\eqref{Eq:half_space_reflection}, we can obtain the theoretical results for the magneto-optical transmission of the sample, which are shown in Fig.~\ref{Fig:total_spectrum_epsilon_S1}(c). 
Note that here we plot $T_{\text{unpolarized}} = (T_{+} + T_{-})/2$ because the experimental data is obtained using an unpolarized light source. 
As demonstrated in the figure, there is a good agreement between our theory and the experimental data.

\subsection{The \thepair\ interband transition}

\begin{figure*} [!]
    \centering
    \includegraphics[width=\textwidth]{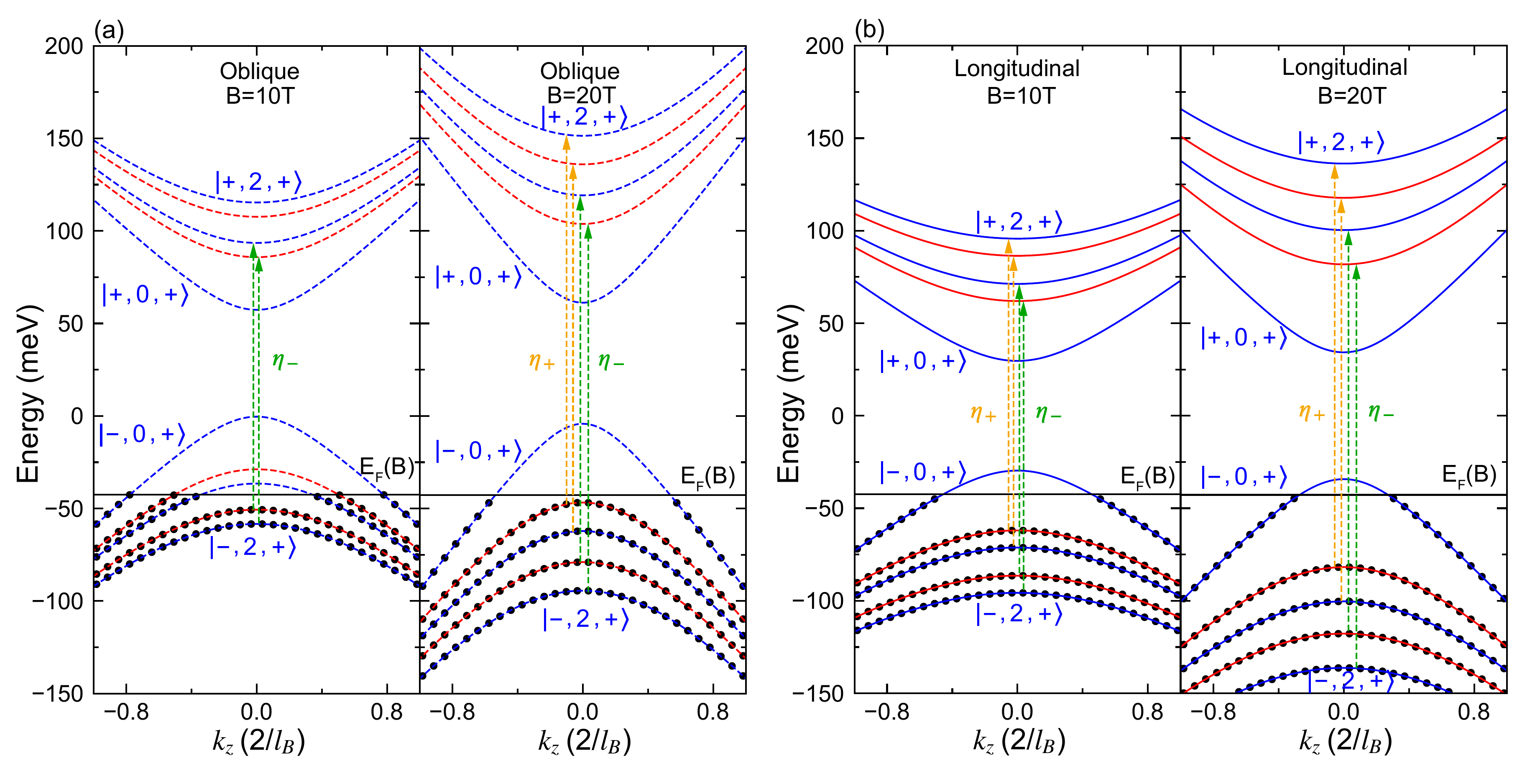}    
    \caption{\label{Fig:onetwo_transitions}
    The theoretical \onetwopair\ interband transition of (a) oblique and (b) longitudinal valleys of the Sample S1 at $B=\SI{10}{T}$ and $B=\SI{20}{T}$.
    Dashed and solid lines show $k_z$ dispersion of the Landau levels of oblique and longitudinal valleys, respectively. 
    The blue (red) lines illustrate the Landau level with the index $\xi=+1$ ($-1$). 
    The black solid lines illustrate the Fermi level, and black dots correspond to the occupied electron states.
    The Landau levels from top to bottom in (a)-(b) correspond to the following states:
    $\ket{+, 2, +}$, $\ket{+, 2, -}$, $\ket{+, 1, +}$, $\ket{+, 1, -}$, $\ket{+, 0, +}$,
    $\ket{-, 0, +}$, $\ket{-, 1, -}$, $\ket{-, 1, +}$, $\ket{-, 2, -}$, $\ket{-, 2, +}$.
    The dashed green lines label the interband transition of the left-handed circularly polarized light $\eta_-: \ket{-, 2, \pm} \rightarrow \ket{+, 1, \mp}$ at $k_z=0$.
    The dashed orange lines label the interband transition of the right-handed circularly polarized light $\eta_+: \ket{-, 1, \pm} \rightarrow \ket{+, 2, \mp}$ at $k_z=0$.  
    }
\end{figure*}

Having achieved a general agreement between the theory and the experimental data, we now focus on the \thepair\ interband transition, which is one of the most interesting features in our data. 
A dashed box in Fig.~\ref{Fig:total_spectrum_epsilon_S1}(c) highlights this part of the data. 

First, this part of the data mainly arises from the group of four transitions labeled as \onetwopair, as defined in Eq.~\eqref{Eq:interband_transition}.
Second, note that there are two dips inside the dashed box, the left of which belongs to the oblique valley and the other to the longitudinal valley. 
Most interestingly, the two dips initially have a similar magnitude at $B = \SI{15}{T}$ but become highly uneven at $B = \SI{35}{T}$. 
This is surprising because one might expect the magnitude of the oblique valley dip to be three times that of the longitudinal valley dip. After all, the material has three oblique valleys and only one longitudinal valley. 

To analyze this unexpected feature, 
note that the different populations in the two valleys can have an impact on the absorption strength of the \onetwopair\ interband transition. 
As shown in Fig.~\ref{Fig:Fermi_level}(a), 
the \onetwopair\ interband transition of the longitudinal valley saturates at $B = \SI{4.2}{T}$, 
whereas the oblique valleys do not achieve that until $B = \SI{18.9}{T}$. 
To further understand those details, we consider the allowed interband transitions of circularly polarized light.
The selection rules for the absorption process are mainly determined by the following matrix elements of the velocity operator $\hat{v}_{\tau=\pm}$,
\begin{align}\label{Eq:interband_matrix_elem}
    &\mel{\lambda',N',\xi',k_z'}{\hat{v}_{\tau=\pm}}{\lambda,N,\xi,k_z} \notg
    =& \delta_{k_z', k_z} \delta_{N', N+\tau} \delta_{\lambda', 1} \delta_{\lambda, -1} (Q_1 + Q_2), 
\end{align}
where 
\begin{align}   \label{Eq:Q_matrix_elem}
    Q_1 &= \frac{\hbar}{l_B m}[(C'C - A'A)\sqrt{N - \frac{1}{2} + \frac{1}{2}\tau} \notg
    &+ (D'D - B'B)\sqrt{N + \frac{1}{2} + \frac{1}{2}\tau}], \\
    Q_2 &= \frac{v}{\sqrt{2}}\qty[(1+\tau)(A'D+C'B) + (1-\tau)(B'C+D'A)],    \notag
\end{align}
and $\tau=+1$ ($-1$) labels the right-handed (left-handed) circularly polarized light.
The general form of eigenstates $\ket{\lambda,N,\xi,k_z}=[A,B,C,D]$, and $A,B,C,D$ are functions of variables $\qty(\lambda,N,\xi,k_z)$.

Because the transition dips primarily arise from $k_z \approx 0$, we will then focus on that part.
Specifically,
for $k_z=0$, the eigenstates take the form
\begin{align} \label{Eq:kz=0_eigenstates}
    \ket{+,N,+} = [A,0,0,D], \notg
    \ket{+,N,-} = [0,B,C,0], \notg
    \ket{-,N,+} = [0,B,C,0], \notg
    \ket{-,N,-} = [A,0,0,D].
\end{align}
By combining the Eq.~\eqref{Eq:interband_transition}, \eqref{Eq:interband_matrix_elem}$-$\eqref{Eq:kz=0_eigenstates}, 
we can identify the allowed interband transitions of the right-handed circularly polarized light ($\eta_+$) and the left-handed circularly polarized light ($\eta_-$) as shown in Fig.~\ref{Fig:onetwo_transitions}.
These transitions correspond to 
\begin{align} \label{Eq:lr_interband_transition}
    \eta_+ &: \ket{\lambda = -, N, \xi = \pm} \rightarrow \ket{\lambda = +, N+1, \xi = \mp} ,  \notg
    \eta_- &: \ket{\lambda = -, N+1, \xi = \pm} \rightarrow \ket{\lambda = +, N, \xi = \mp}.
\end{align}
However, this constraint is generally not satisfied for $k_z\neq0$.
Let us now proceed to examine the \onetwopair\ interband transitions at $k_z=0$ before and after the occupation of $\ket{-,1,\pm}$ of the oblique valley.
As shown in Fig.~\ref{Fig:onetwo_transitions}(a), when $\ket{-,1,\pm}$ is unoccupied by electrons, only $\eta_- : \ket{-,2,\pm} \rightarrow \ket{+,1,\mp}$ transitions are allowed.
After the Fermi level moves above the $\ket{-,1,\pm}$ state at $B = \SI{20}{T}$, all four transitions $\eta_\pm$ in Eq.~\eqref{Eq:lr_interband_transition} become allowed. 
This will increase the absorption strength of the \onetwopair\ interband transition for oblique valleys. 
A similar analysis could be conducted at $B=\SI{15}{T}$ where the allowed interband transitions for the oblique valleys are $\eta_- : \ket{-,2,\pm} \rightarrow \ket{+,1,\mp}$  and $\eta_+ : \ket{-,1,+} \rightarrow \ket{+,2,-}$.
While the longitudinal valley has already reached the quantum limit at $B = \SI{4.2}{T}$, 
the absorption strength of the \onetwopair\ interband transition for the longitudinal valley is expected to remain unchanged as shown in Fig.~\ref{Fig:onetwo_transitions}(b).
In summary, the very different populations in the two valleys (see Table~\ref{Tab:parameters_S1})
can cause the absorption strength ratios to deviate from $3:1$.

\begin{figure}[t]
    \centering
    \includegraphics[width=\columnwidth]{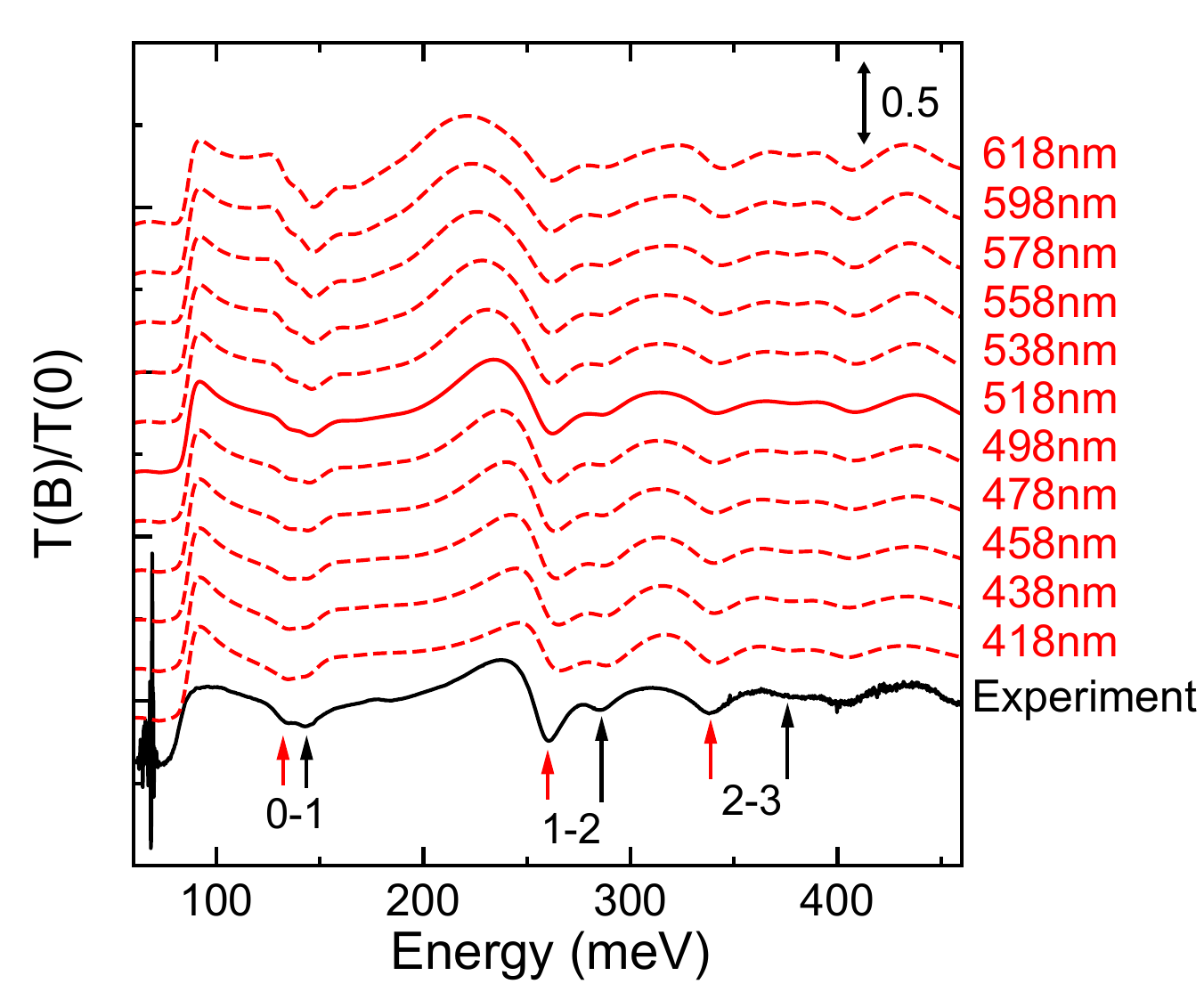}
    \caption{
    \label{Fig:diff_thickness_35T_S1}
    Theoretical transmission spectra at $B=\SI{35}{T}$ with varying sample thicknesses (red dashed or solid lines). 
    The red solid line represents the thickness of the sample.
    The experimental data of sample S1 is represented by the black solid line for reference.}
\end{figure}

However, the above explanation still does not capture a puzzle in the experimental data: above $B = \SI{18.9}{T}$, the \onetwopair\ interband transitions of the oblique valley are already saturated. 
Therefore, the absorption strength ratio of the \onetwopair\ interband transition should not vary significantly after this point. 
Contrarily, the ratio appears to increase even further from \SI{20}{T} to \SI{35}{T}, as shown in Fig.~\ref{fig:3}(c) and Fig.~\ref{Fig:total_spectrum_epsilon_S1}(c).
This proves that the movement of the Fermi energy is not the sole reason for the anomalous absorption strength ratio observed in the experiment.
It turns out that the thickness $d$ of the sample also plays an important role in determining the pattern of the absorption strength ratio. 
In Fig.~\ref{Fig:diff_thickness_35T_S1}, we present the transmission spectra as a function of thickness at $B=\SI{35}{T}$. Notably, a small change in the thickness of about 20\% remarkably alters the shape of the spectra near the \zeroonepair\ and \onetwopair\ interband transition. 
As discussed in Ref.~\cite{palikInfraredMicrowaveMagnetoplasma1970}, the transmission will display an oscillatory behavior as a function of the film thickness $d$ and the frequency $\omega$, especially in the lossless limit ($n \gg k$), which is where our experiment was performed. 
This phenomenon is known as multiple-beam interference or Fabry-P\'{e}rot interference.

Following this observation, it is instructive to compare the slab transmission formula in Eq.~\eqref{Eq:slab_transmission} with the Beer-Lambert law $T = e^{-A d}$, which has been widely used to explain the transmission data in bulk semiconductors. 
We note that the Beer-Lambert law only applies to thick samples, while Eq.~\eqref{Eq:slab_transmission} is much more general. 
To better elucidate their differences, we present a simplified version of Eq.~\eqref{Eq:slab_transmission},
\begin{equation}
    T = \frac{(1 - \qty|r|^2)^2}{e^{A d} -2 \qty|r|^2 \cos(2 \alpha d)}, 
\end{equation}
which is obtained by noting that $\bar{n} \gg k$, $\qty|r|\approx 0.67$, $\phi_r\approx-\pi$ in our sample (see Appendix~\ref{Appendix:OpticalConstants} for more details). 
In our sample, we have $Ad \in \qty(0.04, 1.30)$, and thus, we cannot simply apply the Beer-Lambert law. 
Instead, the factor $\cos(2\alpha d)$ plays an important role, leading to a strong modulation between the longitudinal and oblique valleys. 
This is why a slight change in sample thickness could yield an appreciable variation in the absorption strength ratio between these two valleys.
Such a feature is a direct manifestation of field-dependent multiple-beam interference. 
Note that the importance of multiple-beam interference is also emphasized in the literature in the $B=0$ case~\cite{klannFastRecombinationProcesses1995}.
The Beer-Lambert law is recovered only in the limit $Ad\gg1$, or $d \gg A^{-1} = c/(2\omega k)$. 
For example, from the plot of $k$ in our sample shown in Appendix~\ref{Appendix:OpticalConstants}, we estimate that the Beer-Lambert law only applies to samples with thickness $d > \SI{2.0}{\micro\meter}$ in the frequency range of $\hbar\omega > \SI{200}{meV}$. 

To test our theory further, we applied the same calculation to another sample (S2), which has a different composition and thickness. 
The fitting results are also satisfactory and can be found in Appendix \ref{Appendix:sampleS2}. 

\section{Discussion and Conclusion \label{Sec:Conclusion}}

Although such an anomalous absorption strength ratio variation has been observed in other samples, there are a few reasons why it is particularly prominent in our sample. 
First, the contributions from the longitudinal and oblique valleys should be well separated in the transmission spectra so that we can distinguish the contributions from the two types of valleys. 
This condition is usually satisfied before the band inversion of the material \PbSe\ ($x<0.16$) because the Fermi surfaces of the four $L$ valleys have an ellipsoid shape~\cite{krizmanDiracParametersTopological2018, phuphachongMagneto-spectroscopy2017}. 
Therefore, the longitudinal valley and the oblique valleys have very different $v_c$ corresponding to different dips' positions. 
In contrast, after the band inversion of \PbSe\ ($x>0.16$)~\cite{tikuisisLandauLevelSpectroscopy2021}, the Fermi surfaces of the four $L$ valleys become spherical, and hence the differences between the longitudinal and oblique valleys cease to exist~\cite{krizmanDiracParametersTopological2018}. 
As a result, these anomalous absorption strength ratio variations can be difficult to observe after the band inversion~\cite{tikuisisLandauLevelSpectroscopy2021}.

To summarize, in this work, we carried out a detailed experimental and theoretical study of infrared magneto-optical spectroscopy in the Pb$_{1-x}$Sn$_{x}$Se alloy system. 
In particular, we focus on the variations of the absorption strength ratio of the interband transition, which is often neglected in the literature. 
We developed a theoretical model to describe the fine structures in the interband transition, which helped us extract the dielectric function of the sample. 
The fitting between the experiment and theory is satisfactory. 
The analysis allows us to attribute anomalies in the spectral shape of the infrared magneto-optical spectra to an uneven valley population and the impact of the field-dependent multiple-beam interference. 
The different population has been observed and attributed to the thermal strain between the substrate BaF${_2}$ and lead salts in literature~\cite{burkeMagnetophononShubnikovde1978, assafMassiveMassless2016}.
Meanwhile, the field-dependent multiple-beam interference is most pronounced when the material thickness is sufficiently small. 

Our work sheds light on whether the absorption strength can help identify symmetry-breaking patterns in a sample. 
By carrying out a thorough theoretical and experimental analysis, we identify a large carrier density difference between longitudinal and oblique valleys in our sample.
The fact that a single-particle model provides sufficient evidence to explain the anomalous absorption strength ratio leaves little room for spontaneous symmetry breaking in our sample. 
Our work demonstrates that sample thickness could complicate the analysis. Therefore, a careful analysis is needed to understand the spectral shape.

Our model provides a crucial foundation for identifying other fine structures in the interband transition region resulting from different mechanisms. 
However, there are still unanswered questions that require further investigation. 
First, there exists a deviation of $T(B)/T(0)$ between the experimental results and our calculations in the low-frequency regime, such as the $\SI{120}{meV} < \hbar\omega < \SI{190}{meV}$ window for the $B = \SI{20}{T}$ curve in Fig.~\ref{Fig:total_spectrum_epsilon_S1}(c). 
This is likely due to the existence of an absorption edge within this energy window. 
Such a disagreement has also been observed in other experiments~\cite{tikuisisLandauLevelSpectroscopy2021}. 
Meanwhile, it will be interesting to carry out a similar study for Pb$_{1-x}$Sn$_{x}$Te. 
We anticipate that some of the features in our sample will be even more pronounced in those samples because the difference between the $v_c$ in the longitudinal and oblique valleys is much larger.

\section*{Acknowledgement}
X.D. and X.L. are supported by the Research Grants Council of Hong Kong (Grants~No.~CityU~21304720, CityU~11300421, CityU~11304823, and C7012-21G), and City University of Hong Kong (Projects~No.~9610428 and 7005938). 
Y.-T.H. acknowledges support from NSF Grant No. DMR-2238748. 
This research was supported in part by grant NSF PHY-2309135 to the Kavli Institute for Theoretical Physics (KITP).
JW, XL and BAA acknowledge support from National Science Foundation Award No. DMR-1905277 and DMR-2313441. 
A portion of this work was performed at the National High Magnetic Field Laboratory, which is supported by National Science Foundation Cooperative Agreement No. DMR-2128556 and the State of Florida.

\appendix

\section{Details of determination of the carrier density and the valley splitting \label{Appendix:Determining Carrier Density}}

\begin{figure} [b]
    \centering
    \includegraphics[width=\columnwidth]{"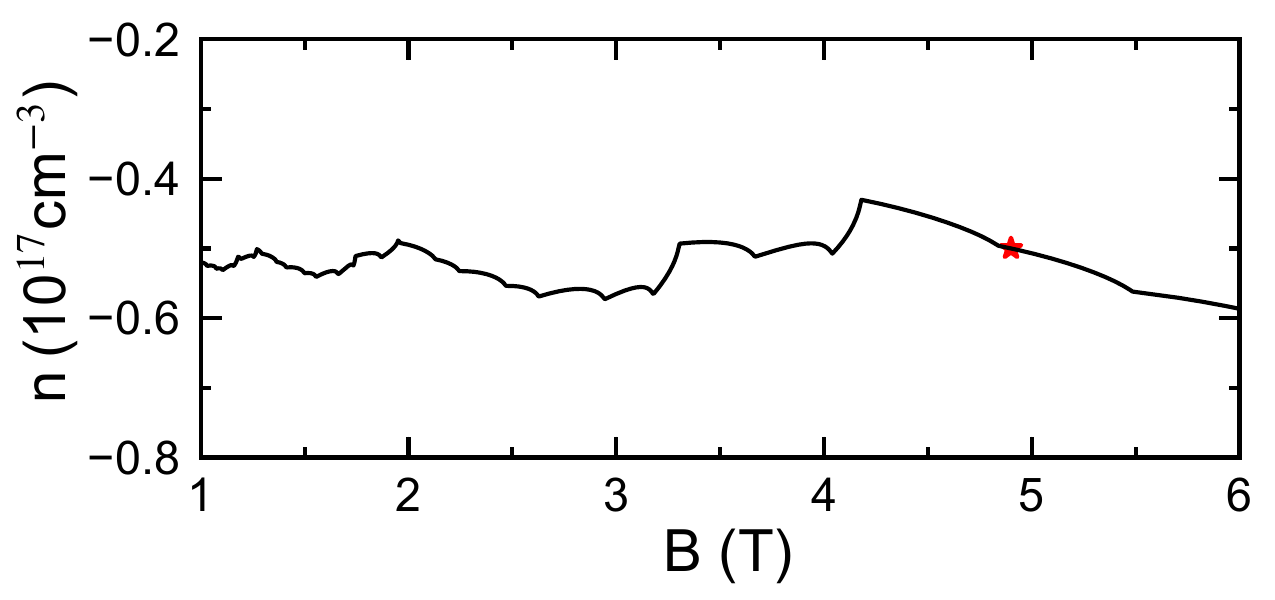"}
    \caption{ The population of the longitudinal valley (solid line) at different magnetic fields. 
    The red star marks the initial guess.
    \label{Fig:self_consistent_carrier_density}}
\end{figure}

In this Appendix, we explain the procedure for determining the carrier density of samples. 
Additionally, we will present the fitting results for the valley splitting.

\begin{table*}[!]
    \centering
    \caption{\label{Tab:footprint_E_F_exp}The onsets of different transitions in the experiment.}
    \begin{threeparttable}
        \begin{tabular}{>{\centering\arraybackslash}p{2cm}|>{\centering\arraybackslash}p{3cm}|>{\centering\arraybackslash}p{4cm}|>{\centering\arraybackslash}p{4cm}|>{\centering\arraybackslash}p{4cm}}
        \hline\hline
            \multirow{2}{*}{Transitions}        & \multicolumn{2}{c|}{Oblique valley}  & \multicolumn{2}{c}{Longitudinal valley}  \\ \cline{2-5}
            \multirow{2}{*}{}& $B_c$(\SI{}{T}) & $n_O(10^{17}/\SI{}{cm}^3)$ & $B_c$(\SI{}{T}) & $n_L(10^{17}/\SI{}{cm}^3)$ \\ \hline
            \zeroonepair\ &     -     &         -            & $3.00$ - $4.00$  &$0.42$ - $0.67$           \\ \cline{2-3} \hline
            \onetwopair\  & $6.00$ - $7.00$ & $2.5$ - $3.1 $                                                  & $1.75$ - $2.50$ &$0.51$ - $0.89$        \\ \cline{2-3} \hline
            \twothreepair\  & $4.75$ - $5.00$ &$3.1$ - $3.4$     & $1.50$ - $2.50$  &$0.74$ - $1.6$       \\ \cline{2-3} \hline \hline
        \end{tabular}
    \end{threeparttable}
\end{table*}

\begin{table*}[!]
    \centering
    \caption{\label{Tab:footprint_E_F_theory}The onsets of different transitions of the theory.}
    \begin{threeparttable}
        \begin{tabular}{>{\centering\arraybackslash}p{2cm}|>{\centering\arraybackslash}p{3cm}|>{\centering\arraybackslash}p{4cm}|>{\centering\arraybackslash}p{4cm}|>{\centering\arraybackslash}p{4cm}}
        \hline\hline
            \multirow{2}{*}{Transitions}        & \multicolumn{2}{c|}{Oblique valley}  & \multicolumn{2}{c}{Longitudinal valley}  \\ \cline{2-5}
            \multirow{2}{*}{}& $B_c$(\SI{}{T}) & $n_O(10^{17}/\SI{}{cm}^3)$ & $B_c$(\SI{}{T}) & $n_L(10^{17}/\SI{}{cm}^3)$ \\ \hline
            \zeroonepair\ &     $12.75$     &  $3.0$                 & $3.30$  &$0.49$           \\ \cline{2-3} \hline
            \onetwopair\  & $7.00$ & $3.1$                                                  & $1.75$ &$0.51$        \\ \cline{2-3} \hline
            \twothreepair\  & $4.84$ &$3.2$     & $1.18$  &$0.51$       \\ \cline{2-3} \hline \hline
        \end{tabular}
    \end{threeparttable}
\end{table*}

\subsection{Challenges in determining the carrier density}
Several factors make it complicated to determine the carrier density in each valley in this experiment. 
First, in the presence of a strong magnetic field, the carrier density in each valley is no longer separately conserved because all four valleys share one single Fermi energy~\cite{bauerMagnetoopticalPropertiesIV1980, burkhardBandpopulationEffectsIntraband1979, adlerModelMagneticEnergy1973}. 
In other words, carriers can transfer between different valleys as the magnetic field strength is varied. 
Second, the exact value of the valley splitting $\Delta_S$ cannot be measured directly from the experiment. 
Third, the experimental data only provides the onset of each transition, which is not sufficient to determine the exact position of the Fermi energy $E_F$ at different magnetic fields. 

To address these challenges, we first determine the carrier density of each valley at a specific magnetic field by identifying the onset of a specific interband transition. 
For example, the onset of the \zeroonepair\ interband transition corresponds to the moment that the system occupies the electron state $\ket{-,1,+,k_z=0}$. 
We can generalize this procedure to other transitions, and obtain the $p$-type carrier density of each valley using
\begin{align}	\label{Eq:carrier_density}
    n_{\text{valley}}(E_F)  &= \frac{g_{L}}{S} \sum_{N,\xi} \int_{-\infty}^{\infty} \qty[1-f(E_{\lambda, N, \xi}(k_z))] \frac{d k_z}{2 \pi} \\
            &= \frac{1}{2 \pi^2 l_B^2} \frac{1}{\hbar v_z} \sum_{N, \xi} \sqrt{E_F^2 - P_{N, \xi}^2} \Theta(-\qty|P_{N, \xi}| - E_F). \notag
\end{align} 
The difference between Eq.~\eqref{Eq:total_carrier_density} and Eq.~\eqref{Eq:carrier_density} is that 
the latter determines how many states are occupied in a single valley even without knowing the exact valley splitting $\Delta_S$ between different valleys. 
The reason is that the carrier density of each valley is determined by the number of occupied states in that valley, which is not affected by the relative shift of Landau levels between different valleys. 
Such a procedure is repeated for different transitions, and the results are summarized in Table~\ref{Tab:footprint_E_F_exp}.
We define the onset magnetic field $B_c$ as the magnetic field at which certain transitions become observable.
We note that $B_c$ of each transition is subject to uncertainty due to the resolution of the experimental data. 
As a result, the carrier density of each valley also carries an uncertainty.

\subsection{Fitting the valley splitting}
\begin{figure}[t]
    \centering
    \includegraphics[width=\columnwidth]{"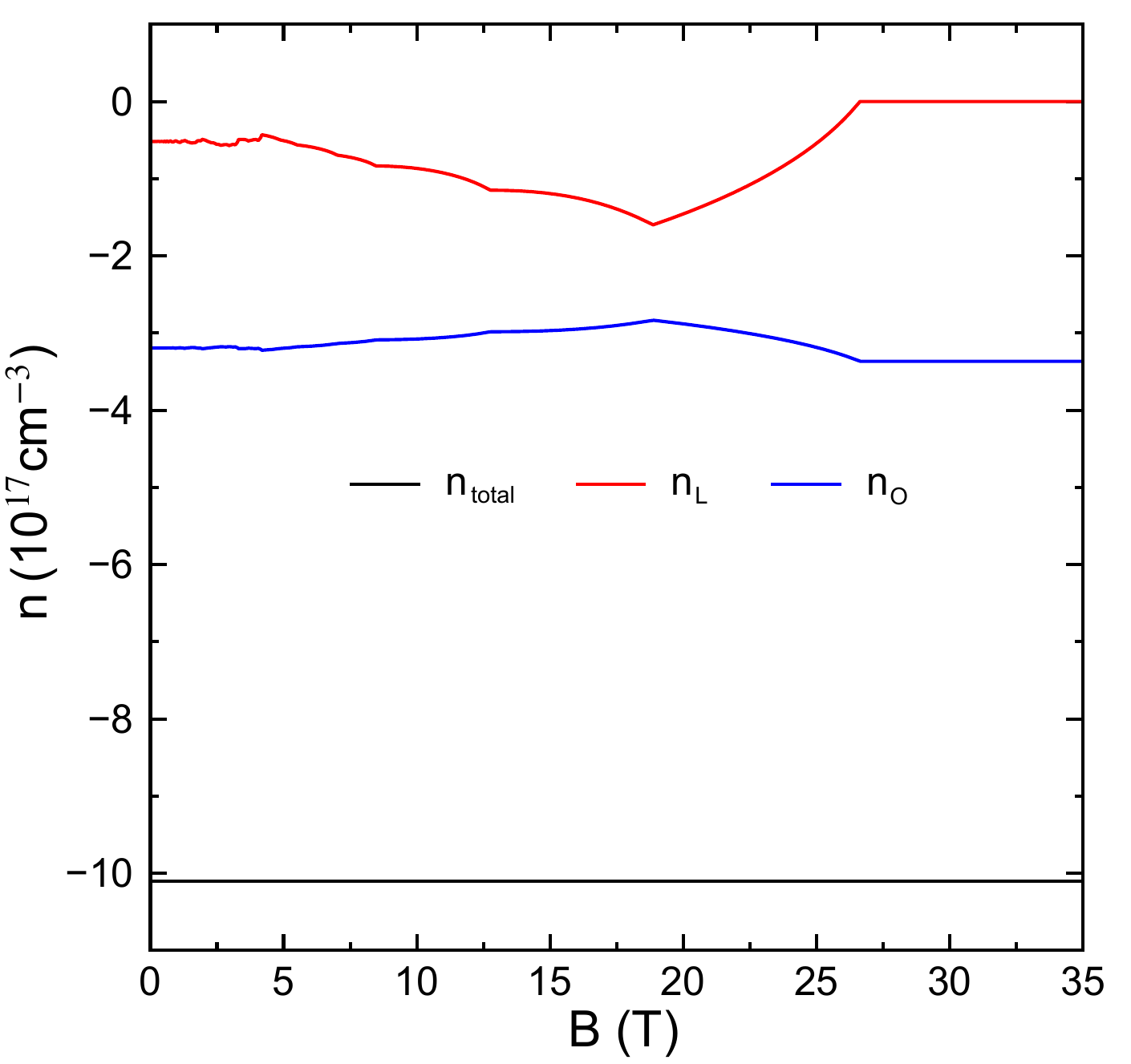"}
    \caption{ The population of different valleys at different magnetic fields. The red and blue solid lines represent the carrier density of the longitudinal and oblique valleys, respectively. 
    The black solid line is the total carrier density of the system. 
    The carrier density of the longitudinal valley vanishes at around $B=\SI{26.6}{T}$.
    \label{Fig:total_carrier_density}}
\end{figure}

From the above discussion, we know that if the onsets of the transitions of two valleys can be identified simultaneously, one can determine the total carrier density of the system uniquely. 
However, this is generally not possible from the experimental data. 
To address this issue, we utilize a fitting procedure. 
Specifically, we start by choosing an initial guess for the carrier densities for longitudinal and oblique valleys and then search for a consistent solution from the following equations of the total particle number conservation: 
\begin{align} \label{Eq:self_consistent_carrier_density}
    &n_O(\Delta_S, B_i) = n_O^i , \notg
    &n_L(\Delta_S, B_i) = n_L^i , \notg
    &n_\text{total} = \frac{g_{L}}{S} \sum_\text{valleys} \sum_{N,\xi} \int_{-\infty}^{\infty} \qty[1-f(E_{\lambda, N, \xi, \Delta_S}(k_z))] \frac{d k_z}{2 \pi}  \notg
        &= 3  n_O^i + n_L^i 
    = 3  n_O(\Delta_S, B) + n_L(\Delta_S, B)    ,
\end{align}
where $n_O^i$ and $n_L^i$ are the initial guesses at a given magnetic field $B_i$, $n_\text{total}$ is the total carrier density of the system. 
$n_O(\Delta_S, B_i)$ and $n_L(\Delta_S, B_i)$ are the carrier densities of the oblique and longitudinal valleys at $B_i$ with a fitted valley splitting $\Delta_S$.
For example, for sample S1, according to the Table~\ref{Tab:footprint_E_F_exp}, 
if we assume that $n_O^i = 3.2 \times 10^{17}/\SI{}{cm}^3$ and $n_L^i = 0.5 \times 10^{17}/\SI{}{cm}^3$ respectively at $B_i=\SI{4.9}{T}$,
one can uniquely determine the valley splitting $\Delta_S$ from above Eq.~\eqref{Eq:carrier_density}$-$\eqref{Eq:self_consistent_carrier_density}, 
as shown in Fig.~\ref{Fig:self_consistent_carrier_density}. 
Although there are different initial carrier densities, from our fitting, we find that if one valley is fixed, e.g., oblique valley, 
then the other valley can only vary within $\pm0.1 \times 10^{17}/\SI{}{cm}^3$. 
Otherwise, no proper solution of $\Delta_S$ that satisfies both Eq.~\eqref{Eq:self_consistent_carrier_density} and Table~\ref{Tab:footprint_E_F_exp} can be found.
From this procedure, we can extract the valley splitting $\Delta_S=\SI{28.50}{meV}$ for sample S1. The valley population between the longitudinal and oblique valleys is also shown in Fig.~\ref{Fig:total_carrier_density}. The carrier density of the longitudinal valley even becomes completely depleted after \SI{26.6}{T}.
By fitting the data, we can theoretically obtain the counterpart of Table~\ref{Tab:footprint_E_F_exp} in Table~\ref{Tab:footprint_E_F_theory}.
Considering the minimum step size $\Delta B=\SI{0.25}{T}$ of the experiment, most theoretical results fall within the range of the experimental data.

\begin{figure} [!]
    \centering
    \includegraphics[width=\columnwidth]{"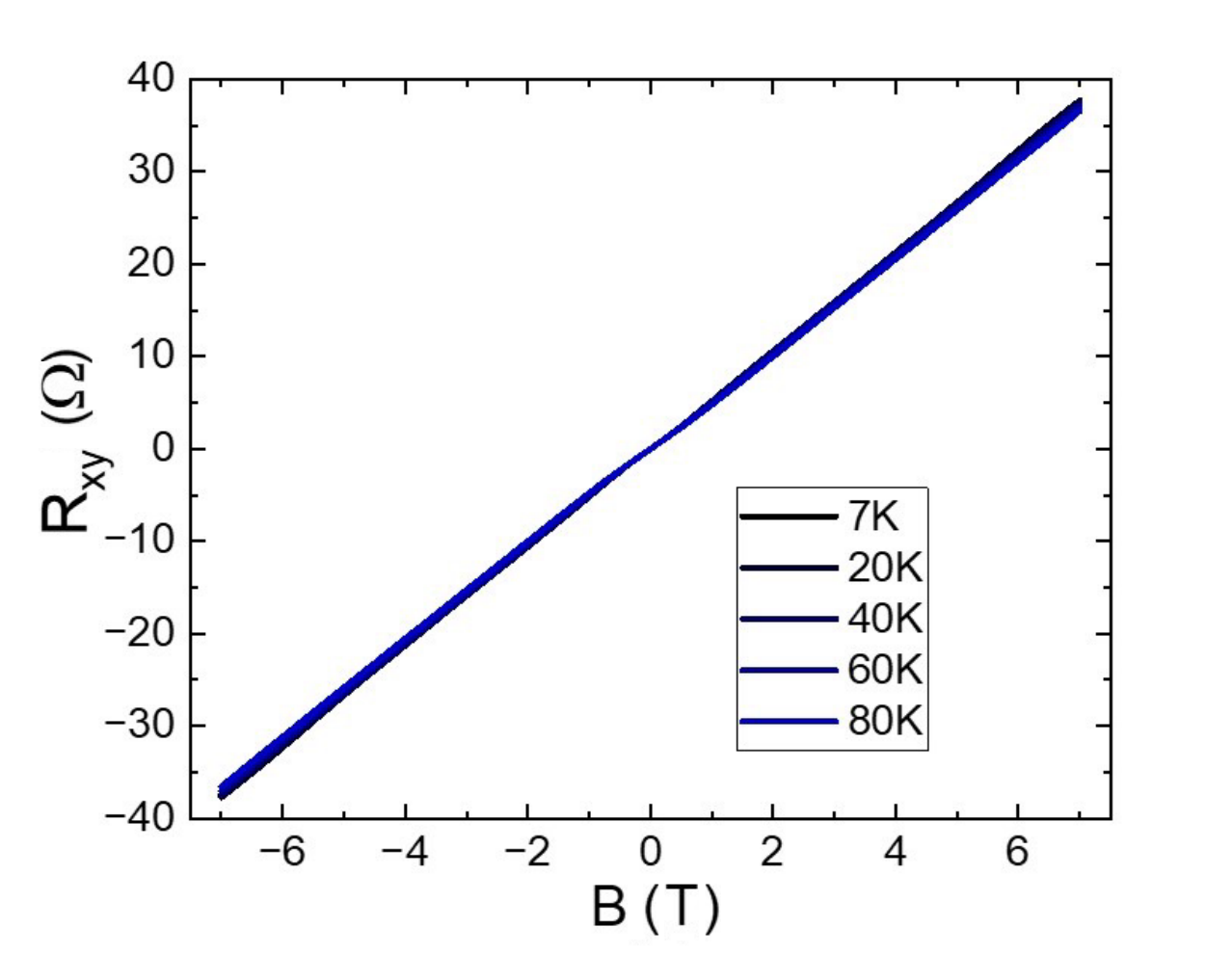"}
    \caption{ The Hall measurement of Sample S2. The positive slope of the Hall resistance indicates the presence of hole carriers of the sample.
    \label{Fig:Hall_measurement}}
\end{figure}

\subsection{Hall measurements for the carrier density}

To identify the type of carrier density, we also performed a Hall measurement 
in Fig.~\ref{Fig:Hall_measurement}. The sample is hole doping due to Sn vacancies. 
From the Hall measurement, we extract the carrier density of the sample S2 to be $p=11 \times 10^{17}/\SI{}{cm^3}$. 
It is close to the one extracted from the infrared magneto-optical spectroscopy $p=7.7 \times 10^{17}/\SI{}{cm^3}$ as indicated in Table~\ref{Tab:parameters_S1}, 
but not totally identical. This kind of difference is also observed in the previous work~\cite{tikuisisLandauLevelSpectroscopy2021}, 
where the difference between two measurements can be as large as a factor of 2.
The difference may be attributed to the inhomogeneity of the sample, 
since Hall measurements are macroscopic, while magneto-optical spectroscopy provides local measurements.

\section{Derivation of the slab transmission \label{Appendix:Slab Transmission}}

In this appendix, we briefly explain how to derive the slab transmission formula in Eq.~\eqref{Eq:slab_transmission}. 

\begin{figure}[!]
    \centering
    \includegraphics[width=\columnwidth]{"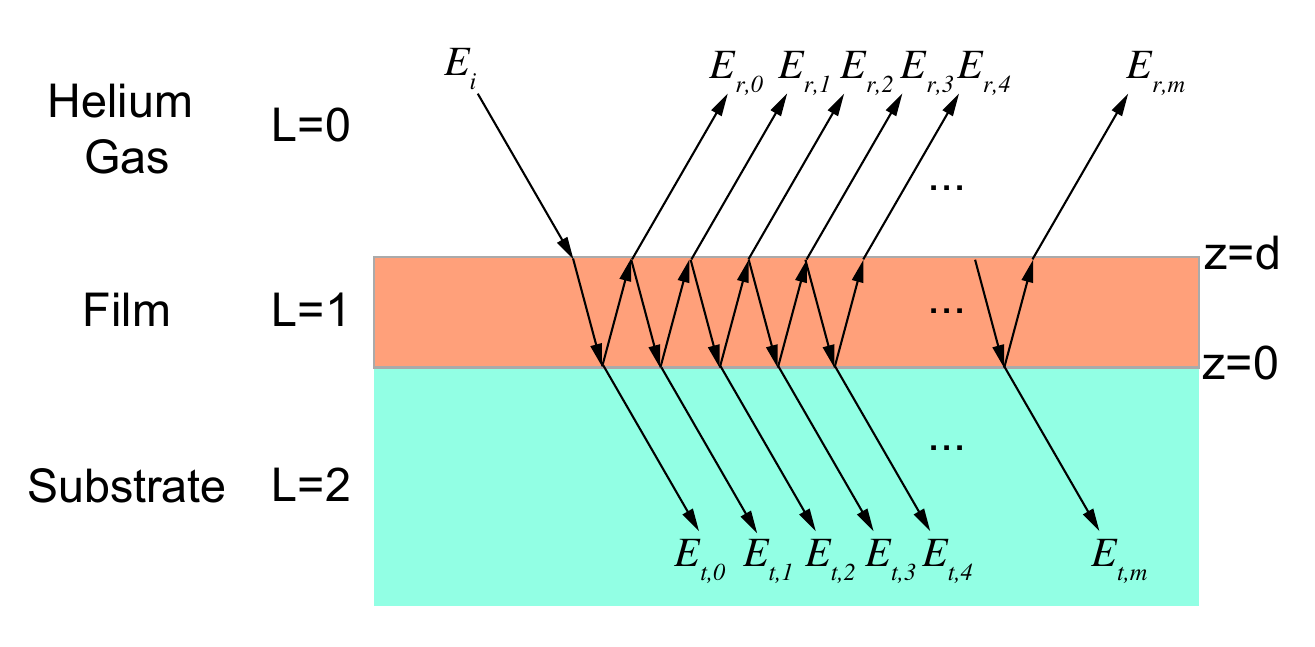"}
    \caption{A sketch of the incidence of a beam with multiple reflections inside a plane-parallel slab of thickness $d$. 
    Here, $L=0,1,2$ labels the index of the layer. 
    In addition, $E_i$, $E_{r,m}$, $E_{t,m}$ label the incident, multiple reflected and transmitted light, respectively.
    \label{Fig:multiple_reflection}}
\end{figure}

We start by considering the geometry shown in Fig.~\ref{Fig:multiple_reflection}, which depicts the general process of a light incident from the helium gas, going through the sample and the substrate, and is eventually received by the bolometer behind the substrate. 
Under this geometry, the complex transmission coefficient $\tau$ can be obtained by summing up the contributions from the incident beam's multiple reflections in the film, which yields 
\begin{align}
	\tau 	&= \frac{\sum_{m=0}^{\infty} E_{t,m}}{E_i} =  t_{01} t_{12} e^{i \delta_1} \sum_{m=0}^{\infty} (r_{12} r_{10} e^{2 i \delta_2})^m  \notg
		&=  \frac{t_{01} t_{12} e^{i \delta_1}}{1 - r_{12} r_{10} e^{2 i \delta_2}}. 
\end{align}
In the above equation, $r_{jk}$ and $t_{jk}$ are the Fresnel coefficients, and $j$ and $k$ label the layer indices of the media shown in the figure. 
$\delta_1$ is the phase accumulated by the incident light when it goes through the film the first time, and $2\delta_2$ is the phase difference between two adjacent beams.
In the normal incident case, $\delta_1 = \delta_2 = \delta$ is the phase accumulated by the incident light when it goes through the film once. 
The phase is given by $ \delta = q d = (\alpha  + i A/2)d$, where $A$ and $\alpha$ were introduced in Eq.~\eqref{Eq:slab_transmission}. 
In addition, the Fresnel coefficients become
\begin{align}
	r_{jk} = \frac{\sqrt{\varepsilon_j} - \sqrt{\varepsilon_k}}{\sqrt{\varepsilon_j} + \sqrt{\varepsilon_k}}, \ 
	t_{jk} = \frac{2 \sqrt{\varepsilon_j}}{\sqrt{\varepsilon_j} + \sqrt{\varepsilon_k}} ,
\end{align}
 In the absorbing media, the complex refractive index is defined as $\sqrt{\varepsilon} = \bar{n} + i k$. 
 Note that the refractive index of the substrate BaF$_2$ in the far infrared regime varies from $1.01$ to $1.46$~\cite{thomasBariumFluorideBaF21997}, which is much smaller than that of \PbSe. Meanwhile, the refractive index of the helium gas is also close to the vacuum.
Therefore, it is reasonable to simplify our model by assuming that the film is surrounded by the vacuum on both sides~\cite{palikInfraredMicrowaveMagnetoplasma1970}. 
 Notice that now $r_{kj} = - r_{jk}$, $r_{01} = -r_{12} = r$, $t_{12} = 1 - r$, $t_{01} = 1 + r$ we then obtain the complex transmission as
\begin{align}
	\tau 	= \frac{t_{01} t_{12} e^{i \delta}}{1 + r_{12} r_{01} e^{2 i \delta}} 
	= \frac{(1-r^2) e^{i \delta}}{1 - r^2 e^{2 i \delta}} .
\end{align}
By substituting the relations $r = \qty|r| e^{i \phi_r}$, we finally obtain the expression for the slab transmission as~\cite{palikInfraredMicrowaveMagnetoplasma1970}
\begin{align}  
    T &= \tau \tau^*  \notg
    &= e^{-Ad} \frac{(1 - \qty|r|^2)^2 + 4\qty|r|^2\sin^2(\phi_r)}{(1 - \qty|r|^2 e^{-A d})^2 + 4\qty|r|^2 e^{-A d} \sin^2(\alpha d + \phi_r)},\notag
\end{align}
which is exactly Eq.~\eqref{Eq:slab_transmission} in the main text.

\section{Details of the numerical procedure \label{Appendix:Numerics}}

In this appendix, we highlight a few notable details from our numerical calculations.

First, to avoid numerical difficulties when calculating the Landau level at a zero magnetic field ($B = 0$), we approximate $T(0)$ with $T(B = \SI{1}{T})$~\cite{tikuisisLandauLevelSpectroscopy2021}. 
In practice, it is sufficient to keep all Landau levels below a maximum Landau level index $N_\text{max}(B)$, which is taken to include all interband transitions below $\SI{800}{meV}$. 
This approximation has been verified against experimental data.
Moreover, the limit of the $k_z$ integral is capped at $k_{z,\text{max}} = \SI{20}{\per\nano\meter}$.

We now comment on the broadening parameter $\Gamma$ in the Kubo formula for conductivity, Eq.~\eqref{Eq:conductivity}. 
Based on the empirical rule for the interband magneto-optical transitions in graphene~\cite{nedoliukColossalInfraredTerahertz2019, orlitaCarrierScatteringDynamical2011}, 
we choose the frequency-dependent broadening parameter $\Gamma(\hbar\omega) = a + b\hbar\omega$~\cite{tikuisisLandauLevelSpectroscopy2021} to fit the experimental data. 

In principle, the broadening parameter $\Gamma$ is the self-energy~\cite{andoElectronicPropertiesTwodimensional1982}, which is a function of the energy of the photon $\hbar\omega$ and the wave vector $\vb*{k}$.
We apply the linear approximation to fit the broadening parameters, given by $\Gamma(\hbar \omega)= \SI{2}{meV} + 0.027 \hbar\omega$.
Note that $\hbar \omega$ is written in the unit of meV.

\begin{figure} [!]
    \centering
    \includegraphics[width=\columnwidth]{"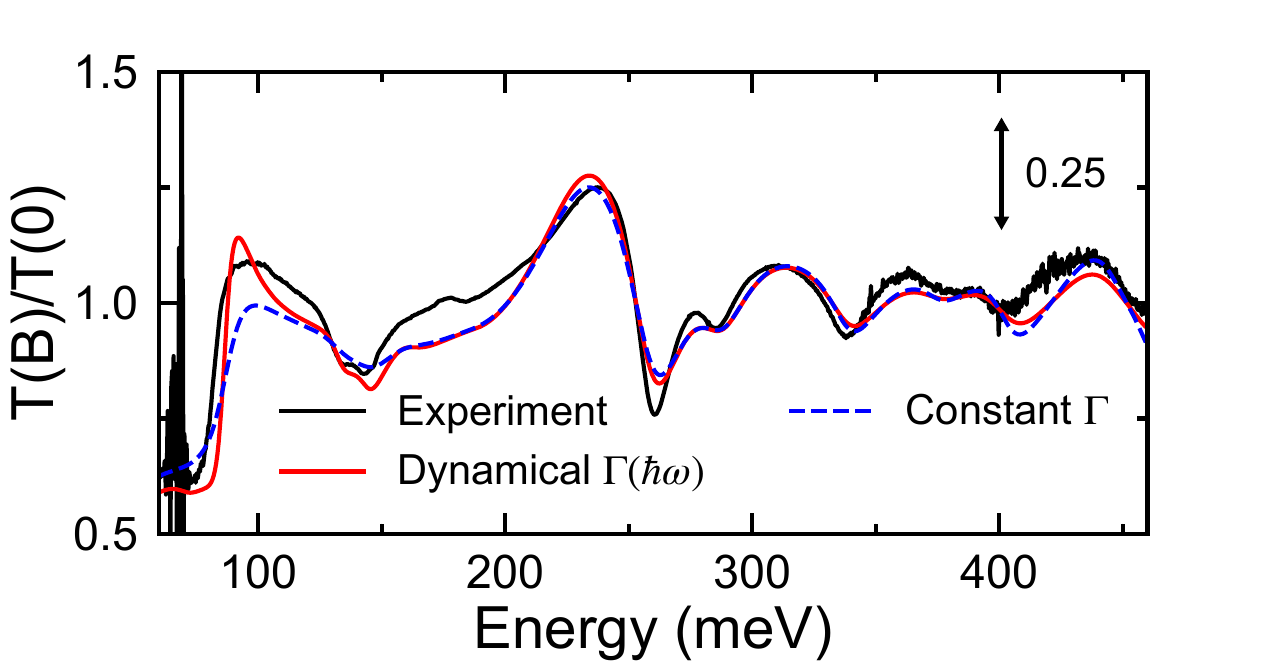"}
    \caption{Differences between a constant and a dynamical broadening parameter at $B=\SI{35}{T}$.  
    The black line illustrates the experimental data of Sample S1. 
    The red line is the theoretical transmission spectra obtained with a dynamical broadening parameter $\Gamma(\hbar \omega)=\SI{2}{meV}+0.027\hbar \omega$. 
    The blue dashed line represents the theoretical transmission spectra obtained with a constant broadening parameter $\Gamma=\SI{10}{meV}$. 
    \label{Fig:constant_broadening_S1}}
\end{figure}

In Fig.~\ref{Fig:constant_broadening_S1}, we compare the results in sample S1 using a constant broadening parameter $\Gamma=\SI{10}{meV}$ with those obtained with the full dynamical broadening parameter $\Gamma(\hbar \omega)= \SI{2}{meV} + 0.027 \hbar\omega$. 
The constant broadening $\Gamma=\SI{10}{meV}$ fit the data well for photon energies above $\hbar\omega=\SI{300}{meV}$, but smears out the dips of \zeroonepair\ interband transition between $\hbar\omega=\SI{100}{meV}$ and $\hbar\omega=\SI{200}{meV}$.
It is evident that the dynamical broadening parameter fits better with the experimental data across the entire spectra. 
Therefore, we used the full dynamical broadening parameter in the results shown in the main text. 
Meanwhile, we find that the same form of $\Gamma$ also works well for the other sample (S2), where we found that $\Gamma(\hbar \omega)=\SI{1}{meV}+0.027\hbar \omega$. 

\section{Additional discussions on the optical coefficients for Sample S1 \label{Appendix:OpticalConstants}}

\begin{figure}[!]
    \centering
    \includegraphics[width=\columnwidth]{"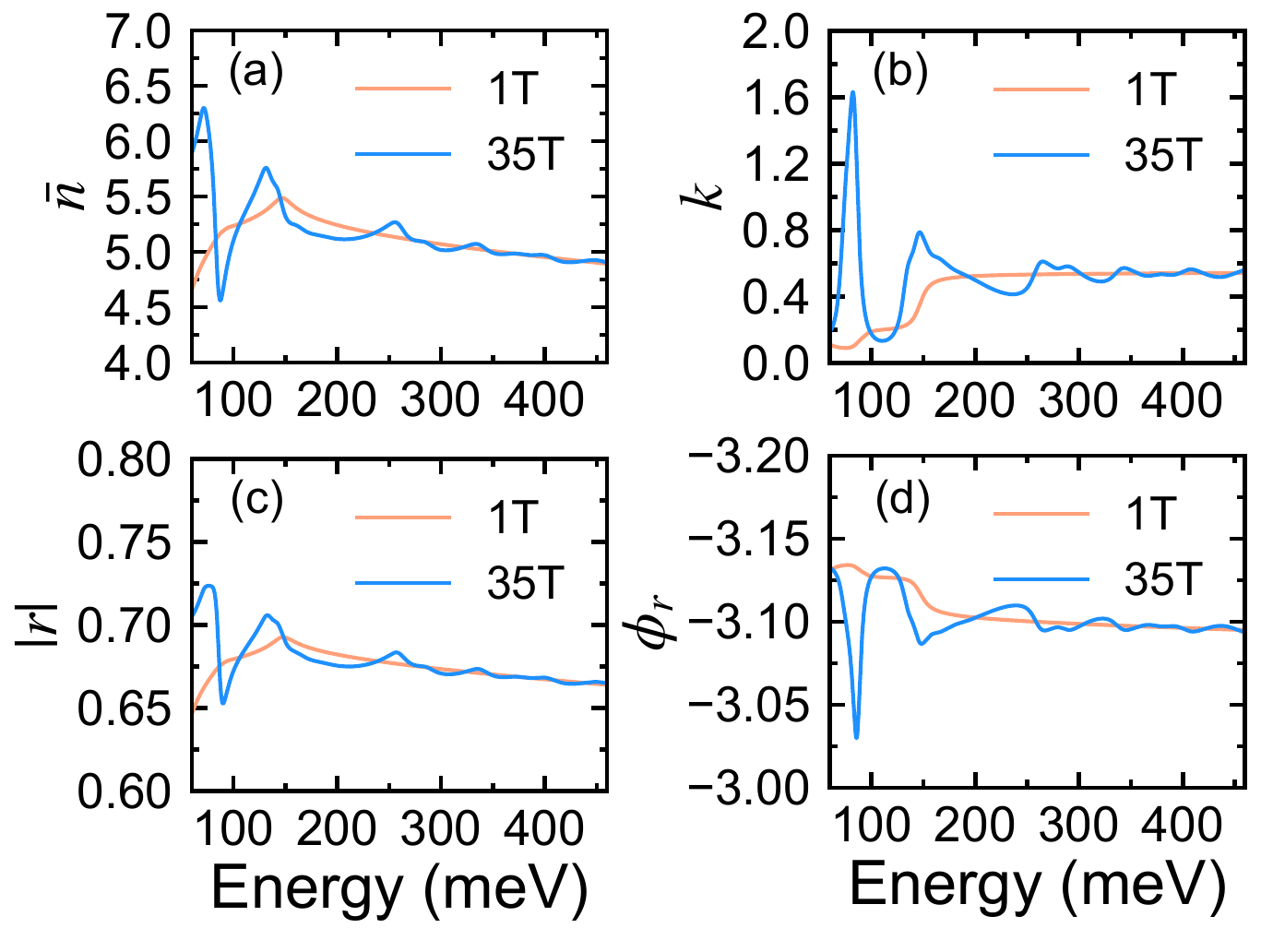"}    
    \caption{
    In this figure, we present theoretically calculated (a) refractive index $\bar{n}$, (b) extinction coefficient $k$, (c) amplitude, and (d) phase of the half-space coefficient $r$ of the sample S1. 
    These results are obtained by averaging over left-handed and right-handed circularly polarized light.  
    \label{Fig:optical_constants_S1}}
\end{figure}

\begin{figure}[!]
    \centering
    \includegraphics[width=0.9\columnwidth]{"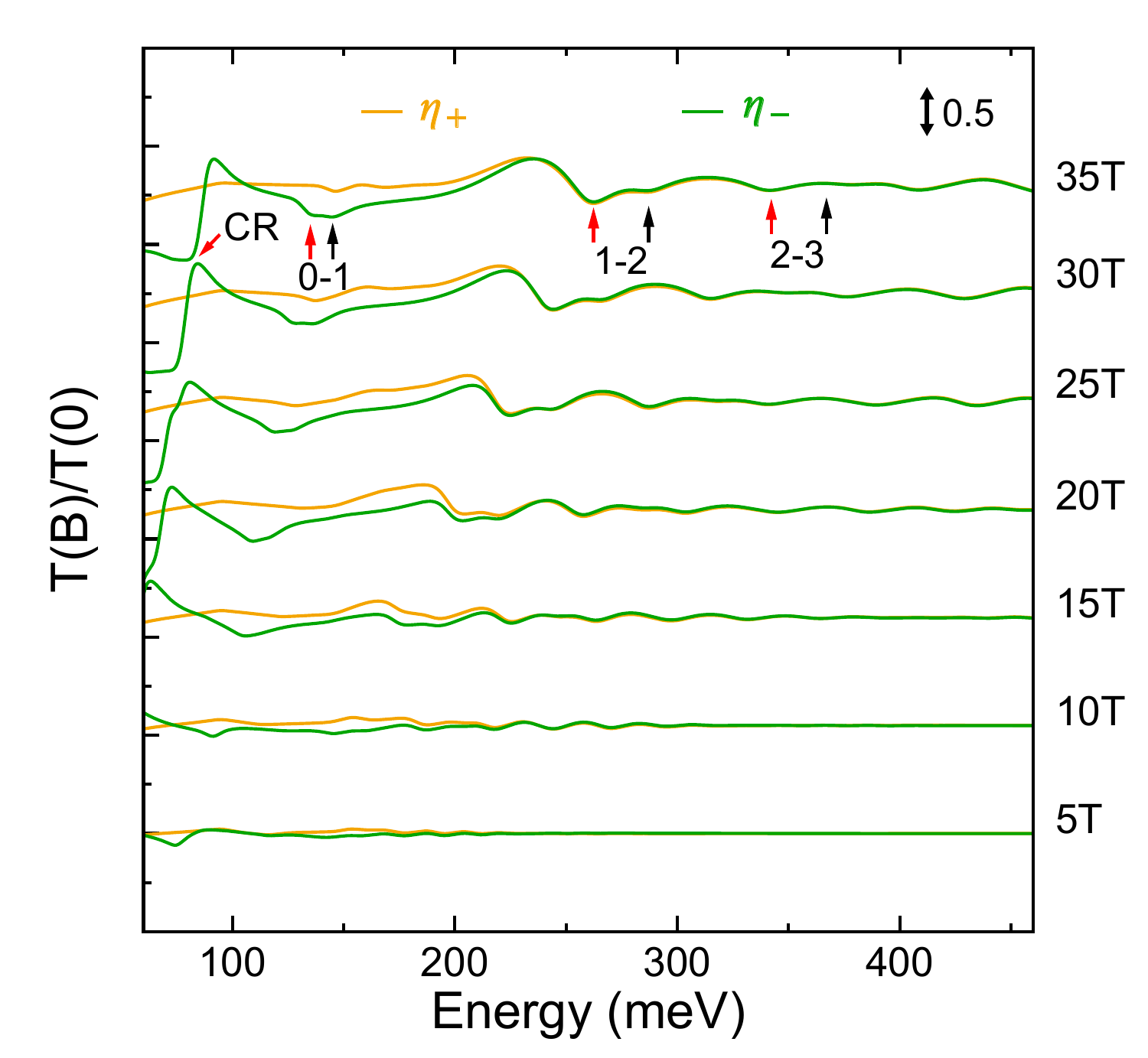"}
    \caption{The theoretical transmission spectra of right-handed ($\eta_+$) and left-handed ($\eta_-$) circularly polarized light of the sample S1.
    \label{Fig:lr_cmp}}
\end{figure}

\begin{figure}[!]
    \centering
    \includegraphics[width=0.9\columnwidth]{"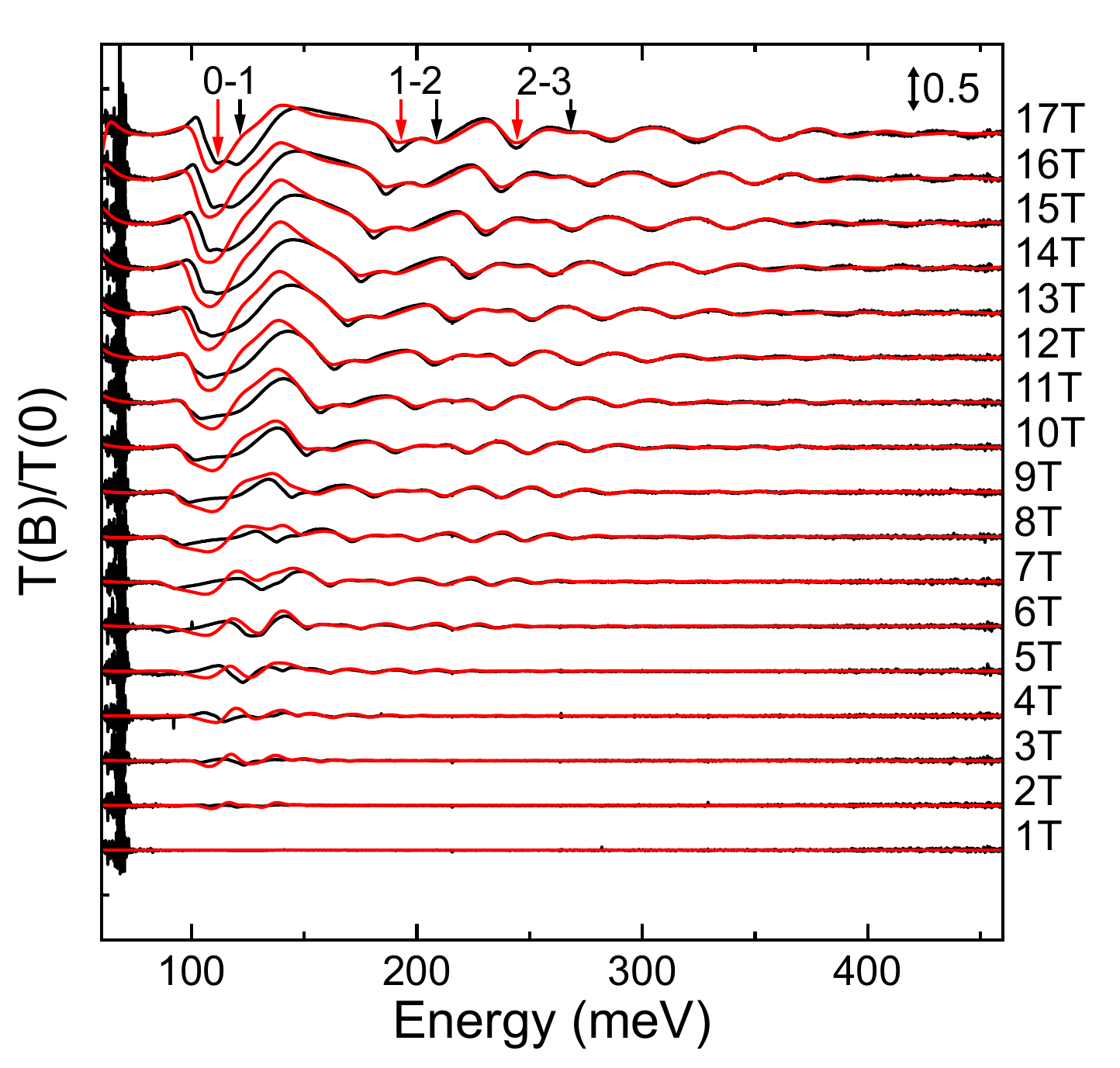"}
    \caption{The infrared magneto-optical spectra of the sample  S2. 
    The black and red lines correspond to the experimental data and theoretical results, respectively. 
    \label{Fig:total_spectrum_S2}}
\end{figure}

Apart from the dielectric function shown in the main text, it is instructive to study other optical constants, including the refractive index $\bar{n}$, the optical extinction coefficient $k$, and the half-space reflection coefficient $r$. 
These quantities are plotted in Fig.~\ref{Fig:optical_constants_S1}. 
We can see that these optical constants satisfy the following relations: $\bar{n} \gg k$, $\abs{r} \approx 0.67$ and $\phi_r \approx -\pi$. 

In addition, at $B=\SI{1}{T}$, both $\bar{n}$ and $\abs{r}$ show two peaks at small photon energies, 
corresponding to the absorption edge of the longitudinal and oblique valleys, respectively. 
The fact that the dispersion of the refractive index will display peaks at the absorption edge has been reported in the previous Ref.~\cite{sternDispersionIndexRefraction1964}.
Meanwhile, the extinction coefficient $k$ displays two steep increases at the same position and keeps flat over photon energy $\SI{200}{meV} < E <\SI{460}{meV}$.
In general, the refractive index $\bar{n}$ and the extinction coefficient $k$ respectively display the same behavior as 
the real and imaginary part of the dielectric function $\varepsilon$ based on Eq.~\eqref{Eq:refractive_index}, and a good agreement of these quantities has been achieved comparing with the literature~\cite{shaniCalculationRefractiveIndexes1985, suzukiOpticalPropertiesPbSe1995, cardonaOpticalPropertiesBand1964b}.

In Fig.~\ref{Fig:lr_cmp}, we compare the transmission spectra of right-handed circularly polarized light ($\eta_+$) and left-handed circularly polarized light ($\eta_-$). 
The spectra exhibit significant differences between the two polarizations.
First, for CR or the \zeroonepair\ intraband transition, the material absorbs more left-handed circularly polarized light $\eta_-$ than right-handed circularly polarized light $\eta_+$.
To understand this behavior, we examine the dipole transition matrix element given by
\begin{align}\label{Eq:intraband_matrix_elem}
    &\mel{\lambda',N',\xi',k_z'}{\hat{v}_{\tau=\pm}}{\lambda,N,\xi,k_z} \notg
    =& \delta_{k_z', k_z} \delta_{N', N+\tau} \delta_{\lambda', -1} \delta_{\lambda, -1} (Q_1 + Q_2),
\end{align}
where $Q_1$ and $Q_2$ are defined in Eq.~\eqref{Eq:Q_matrix_elem} in the main text. 
Based on this equation, we can determine the selection rule for the intraband transition,
\begin{align} \label{Eq:lr_intraband_transition}
    \eta_+ &: \ket{\lambda = -, N, \xi = \pm} \rightarrow \ket{\lambda = -, N+1, \xi = \pm} ,  \notg
    \eta_- &: \ket{\lambda = -, N+1, \xi = \pm} \rightarrow \ket{\lambda = -, N, \xi = \pm}.
\end{align}
When $\ket{-,0,+}$ is the only unoccupied state, only the $\eta_-$ intraband transition $\ket{-,1,+} \rightarrow \ket{-,0,+}$ is allowed at $k_z=0$. This explains why the absorption strength is larger for $\eta_-$ compared to $\eta_+$.
Moreover, because the Landau levels of the valence bands of the longitudinal valley are completely occupied above $B=\SI{26.6}{T}$,  we predict that the CR for that valley is absent in the spectra at $B=\SI{30}{T}$ and $B=\SI{35}{T}$, 
although the experimental data does not provide enough details of this transition in the whole spectra because of the noise. 
Second, we find an enhancement of the \onetwopair\ interband transition of the oblique valley for $\eta_+$ between $B=\SI{10}{T}$ and $B=\SI{20}{T}$ in Fig.~\ref{Fig:lr_cmp}. 
The reason is that, as $B$ increase, the $\ket{-,1,\pm}$ states becomes occupied, and thus enabling more \onetwopair\ interband $\eta_+$ process. 
Finally, for the \zeroonepair\ interband transition, 
the absorption strength of $\eta_-$ is significantly larger than that of $\eta_+$. 
This can be understood from the selection rule in Eq.~\eqref{Eq:lr_interband_transition}, which shows that the \zeroonepair\ interband transition of $\eta_-$ for both valleys can saturate at a lower magnetic field than that of $\eta_+$.

\section{Fitting results for Sample S2 \label{Appendix:sampleS2}} 

Here, we comment on the experimental data in the other sample S2, which has a Sn concentration of $x=0.10$ and a thickness of $\SI{990}{nm}$. 
On the theoretical side, we applied the same procedure to this sample. 
The model parameters are shown in Table~\ref{Tab:parameters_S1}. 
It is notable that the thickness of this sample is almost twice that of the other sample. 
With a slightly different choice of the broadening parameter $\Gamma(\hbar \omega)=\SI{1}{meV}+0.027\hbar \omega$, we can achieve a good agreement with the experimental data for this sample as shown in Fig.~\ref{Fig:total_spectrum_S2}. 
In particular, the anomalous variation of the absorption strength ratios for the \onetwopair\ interband transition is also well captured. 
This agreement demonstrates that our theory is general and can be applied to different samples of this material.

\bibliography{ref9.bib}

\end{document}